\pdfoutput=1 % to force arxiv to use pdflatex
\documentclass[prb,twocolumn]{revtex4-1}

\usepackage{graphicx,color,transparent}
\usepackage{amssymb}
\usepackage{bm}
\usepackage{amsmath,amsfonts,latexsym}

\usepackage{caption}
\usepackage{subcaption}

\usepackage{braket} % For bra-ket notation

\usepackage[colorinlistoftodos,prependcaption,textsize=tiny]{todonotes}

\usepackage{comment}

\usepackage{bm} % allows bold greek letters

\newcommand*{\bk}{\mathbf{k}}

\newcommand*{\cA}{{\cal A}}

\newcommand*{\cD}{{\cal D}}

\newcommand*{\cH}{{\cal H}}

\newcommand*{\cV}{{\cal V}}

\newcommand*{\cO}{{\cal O}}

\newcommand*{\pop}{\psi^{\vphantom{\dagger}}}
\newcommand*{\pdop}{\psi^\dagger}

\newcommand*{\aop}{a^{\vphantom{\dagger}}}
\newcommand*{\adop}{a^\dagger}

\newcommand*{\dop}{d^{\vphantom{\dagger}}}
\newcommand*{\ddop}{d^{\dagger}}

\DeclareMathOperator{\PV}{\mathrm{P}}
\DeclareMathOperator{\sech}{\mathrm{sech}}

\DeclareMathOperator{\mbeq}{\overset{!}{=}}

\DeclareGraphicsExtensions{.pdf}

\newcommand{\hc}{\mathrm{h.c.}}

\newcommand{\dd}[1]{\mathrm{d}#1\,}

\newcommand{\pbyp}[2]{\frac{\partial #1}{\partial #2}}

\begin{document}

\title{Quantum Hydrodynamics in One Dimension beyond the Luttinger Liquid}
\author{Tom Price}
\author{Austen Lamacraft}
\affiliation{TCM Group, Cavendish Laboratory, University of Cambridge, J. J. Thomson Ave., Cambridge CB3 0HE, UK}
\date{\today}
\email{tp294@cam.ac.uk}

\date{\today}

\begin{abstract}
	
  Recent years have seen the development of a rich phenomenology beyond the Luttinger Liquid model of one dimensional quantum fluids, arising from interactions between the elementary phonon excitations. It has been known for some time, however, that the straightforward inclusion of these interactions presents technical difficulties that have necessitated approaches based on refermionization or effective impurity models. 

  In this work we show that the nonlinear extensions of the Luttinger model are tractable in the phonon basis. We present a calculation of the singularities present in the zero temperature dynamical structure factor in the semiclassical limit where the phonon dispersion is strong.

  A unitary transformation decouples interactions between left-- and right--moving phonons, leaving a nonlinear chiral Hamiltonian. At low momenta, this Hamiltonian has a spectrum bounded above and below by thresholds identified with phonon and soliton excitations in the semiclassical limit. The chiral dynamical structure factor therefore has support only in this region, with power law singularities at the thresholds originating in the Anderson orthogonality catastrophe, which we calculate analytically. The dynamical structure factor for the original nonchiral Hamiltonian is a convolution of this chiral correlator with a power law arising from the left--right decoupling.
\end{abstract}

\maketitle

\section{Introduction}

\subsection{The Luttinger Liquid Theory} % (fold)
\label{sub:the_luttinger_liquid_theory}

% subsection the_luttinger_liquid_theory (end)

The Luttinger Liquid (LL) theory \cite{Stone:1994, Giamarchi:2004, Gogolin:2004, Haldane:1981} describes the low energy properties of one dimensional quantum fluids in terms of noninteracting, linearly dispersing density fluctuations. 

In this introductory section we will outline the LL theory %of one dimensional fluids\cite{Stone:1994, Giamarchi:2004, Gogolin:2004, Haldane:1981}
, using the example of the Bose gas\cite{Haldane:1981a}. This will allow us to put the LL theory in the more general context of the hydrodynamic description of quantum fluids, sometimes known as Quantum Hydrodynamics (QHD) \cite{Pitaevskii:2003}. This more general framework allows for both dispersion and interaction of collective excitations, which will be our main concern.

The second quantized Hamiltonian of a gas of bosons of mass $m$ is
\begin{equation}\label{eq:1DBG}
	H_\text{BG} = \int \dd{x\,}\left[\frac{(\partial_{x}\pdop)(\partial_x\pop)}{2m}+\frac{c}{2}\pdop\pdop\pop\pop\right],
\end{equation}
(we set $\hbar =1$), where $c$ governs the strength of short range interactions between the particles, and $\pdop(x)$, $\pop(x)$ are the creation and annihilation operators, obeying the usual commutation relations
\begin{equation}\label{eq:CCR}
	\left[\pop(x),\pdop(y)\right]=\delta(x-y).
\end{equation}
The starting point for the hydrodynamic description is the representation of the field operators in terms of density $\rho(x)$ and phase $\theta(x)$ variables
\begin{equation}\label{eq:DensityPhase}
	\pop(x)=\sqrt{\rho(x)}e^{i\theta(x)}.
\end{equation}
The commutation relations Eq.~\eqref{eq:CCR} are reproduced if $\rho(x)$ and $\theta(x)$ are canonically conjugate
\begin{equation}\label{eq:rhotheta}
	\left[\rho(x),\theta(y)\right]=i\delta(x-y).
\end{equation}
%
%The periodicity of the phase variable is linked to the quantization of particle number, which invites caution in treating $\rho(x)$ as a field taking continuous values. The analysis of Ref.~\onlinecite{Jevicki:1978} however shows that...
In terms of $\rho(x)$ and $\theta(x)$, the Hamiltonian Eq.~\eqref{eq:1DBG} takes the QHD form
\begin{equation}\label{eq:BGQHD}
	H_\text{BG} = \int \dd{x\,} \left[\frac{\left(\partial_{x} \theta\right)\rho \left(\partial_{x} \theta\right)}{2m}+\frac{(\partial_{x} \sqrt{\rho})^{2}}{2m}+\frac{c}{2}\rho^{2}\right].
\end{equation}
An interesting feature of passing to this representation is that the anharmonicity is shifted from the interaction term to the kinetic term.

Dynamics near a of state of uniform density can be described by writing $\rho(x)= n + \varrho(x)$, where $n$ is the mean density. Expanding $H_\text{BG}$ to quadratic order then gives $H_\text{BG}=H_\text{Q}+\ldots$ with
\begin{equation}\label{eq:H2}
	H_\text{Q} = \int \dd{x\,} \left[\frac{n\left(\partial_{x} \theta\right)^{2}}{2m}+\frac{(\partial_{x} \varrho)^{2}}{8mn}+\frac{c}{2}\varrho^{2}\right].	
\end{equation}
This Hamiltonian is quadratic and may be diagonalized by expressing the fields in terms of their Fourier modes
\begin{equation}\label{eq:FourierFields}
\begin{split}
	\varrho(x) = \frac{1}{\sqrt{L}}\sum_p \varrho_p e^{ipx}\\
	\theta(x) = \frac{1}{\sqrt{L}}\sum_p \theta_p e^{ipx},	 
\end{split}
\end{equation}
where $p = 2\pi n/L$, $n=0,\pm 1, \pm 2, \ldots$ for a system of length $L$, and $\varrho_p^\dagger=\varrho_{-p}$, $\theta_p^\dagger=\theta_{-p}$. In terms of the Fourier modes $H_\text{Q}$ takes the form
\begin{equation}\label{eq:FourierH2}
			H_\text{Q} =	\sum_{p\geq 0} \left[\frac{n p^2}{m}|\theta_p|^2 + \left(c + \frac{p^2}{4mn}\right)|\varrho_p|^2 \right].
\end{equation}
Bearing in mind that $\rho_q$ and $\theta_{-q}$ are canonically conjugate, Eq.~\eqref{eq:FourierH2} takes the form of a sum of oscillator Hamiltonians from which we can read off the frequency
\begin{equation}\label{eq:BogFreq}
	\Omega_p^2 = \frac{p^2}{2m}\left(\frac{p^2}{2m}+2cn \right), 
\end{equation}
which is the famous Bogoliubov dispersion relation. At low momentum the Bogoliubov spectrum is linear
\begin{equation}\label{eq:BogLin}
	\Omega_p \sim up + O(p^3),\qquad u^2 = \frac{cn}{m}.  
\end{equation}
Since we are often interested in this limit, it is common to eliminate the term causing the dispersion from the Hamiltonian. One may trace this back to the second term of Eq.~\eqref{eq:BGQHD}, sometimes referred to as the `quantum pressure'. Upon striking it from the Hamiltonian the quadratic Hamiltonian $H_\text{Q}$ becomes the \emph{Luttinger Hamiltonian}
\begin{equation}\label{eq:LuttingerHam}
  H_\text{Q}\to H_{\text{LL}} = \frac{u}{2\pi}\int_0^L \dd{x\,} \left[K(\partial_x\theta)^2 + \frac{1}{K}(\partial_x \phi)^2\right].
\end{equation}
where $\varrho(x)=-\partial_x\phi(x)/\pi$ defines a field $\phi(x)$ with commutation relation
\begin{equation}\label{eq:LLcomm}
  [\theta(x), \partial \phi(y)] = i \pi \delta(x-y),  
\end{equation}
and the dimensionless quantity $K$ --- known as the Luttinger parameter --- is given by
\begin{equation}\label{eq:LuttingerParam}
	K = \frac{\pi n}{mu}. 
\end{equation}
As the above presentation shows, two distinct approximations are involved in describing the Bose gas by the Luttinger Hamiltonian Eq.~\eqref{eq:LuttingerHam}. These are:
\begin{enumerate}
  \item Truncation to a quadratic Hamiltonian. As is clear from Eq.~\eqref{eq:BGQHD}, the expansion around uniform density yields cubic terms in the next order, which will turn out to be the most important.
  \item Dropping the higher derivative (`quantum pressure') term that leads to phonon dispersion.
\end{enumerate}
Usually, these approximations are justified by declaring $H_\text{LL}$ a low-energy effective theory, with the dropped terms being irrelevant perturbations that serve to renormalize the parameters $u$ and $K$.

As should be clear from the above, the QHD Hamiltonian Eq.~\eqref{eq:BGQHD} describes the Bose gas in any dimension. However, in one dimension, the Luttinger Hamiltonian may be used to describe the low energy limit of an \emph{arbitrary} quantum fluid, whose microscopic constituents may in fact be fermions, bosons, or spins. For example, a fermion system may be written in terms of bosons using a Jordan--Wigner transformation, and then the bosons represented as above. 

In this way, the long wavelength part of the operators describing the constituents may be written in terms of the $\phi(x)$ and $\theta(x)$ fields, and since Eq.~\eqref{eq:LuttingerHam} describes a system of non-interacting phonons, the asymptotic behavior of correlations functions may be calculated.
%Quadratic in the fields, thermodynamic quantities such as compressibility, spin susceptibility, etc., may be calculated exactly in terms of the parameters $u$ and $K$. We can then justify 

Despite its apparent limitation to one dimensional problems, the Luttinger Liquid has also been of great importance in understanding two of the great problems of twentieth--century condensed matter physics, the X--ray edge singularity and the Kondo effect (in fact, the bosonization method owes part of its development to the X--ray problem). Both cases may be formulated as scattering problems, allowing a solution by bosonization in a given angular momentum channel \cite{Schotte:1969, Schotte:1970, Giamarchi:2004, Gogolin:2004}.

%To shed some light on the parameters $u$ and $K$ that appear in the Luttinger description, consider the response of the system to a Galilean boost of $P=NJ$, implemented by the zero mode of $\theta \to \theta + Jx/L$. The Hamiltonian \eqref{eq:LuttingerHam} shifts by
%
%\begin{equation}
%  \frac{uK}{2\pi} \frac{J^2}{L},
%\end{equation}
%
%so that the combination $uK$ describes the response to boosts
%
%\begin{equation}\label{eq:HamiltonianBoostResponse}
%  \frac{uK}{\pi} = L \pbyp{^2E}{J^2}.
%\end{equation}
%

%For Galilean invariant systems, the energy must transform by $NJ^2/2m$, and hence $uK = \pi n/m$. 
%\begin{comment}
%	To implement the boost numerically, one employs the twisted the boundary% condition $\psi(L) = e^{iJ}\psi(0)$.
%\end{comment} 
%
%Similarly, $\phi \to \phi - \pi \Delta{N} x/L$ shifts the density by $\Delta{N}/L$. The resulting shift in the Luttinger Hamiltonian, Eqn.~\eqref{eq:LuttingerHam}, fixes the combination $u/K$,

%\begin{comment}
%  \begin{equation}
%    \frac{\pi u}{2K} \frac{(\Delta{N})^2}{L},
%  \end{equation}  
%\end{comment}
%
%\begin{equation}\label{eq:HamiltonianNumberResponse}
%  \frac{\pi u}{K} = L \pbyp{^2E}{N^2}.
%\end{equation}
%
%Calculation of the derivatives for free fermions reveals that $K=1$ and $u$ is the Fermi velocity. However, given a microscopic Hamiltonian lacking in a perturbative expansion, a direct calculation of the Luttinger parameters $u$ and $K$ will typically be out of reach, and they must be pinned down by numerical or experimental evaluation of, e.g., the Eqns~\eqref{eq:HamiltonianBoostResponse},~\eqref{eq:HamiltonianNumberResponse}.

The dynamics of the Luttinger model is conveniently pictured in the chiral basis $\phi_{R/L} = \frac{1}{\sqrt{2\pi}}[\sqrt{K}\theta \mp \frac{1}{\sqrt{K}}\phi]$, with mode expansion
\begin{equation}\label{eq:ChiralCurrentModeExpansion}
  \phi_{R/L} = -\frac{i}{\sqrt{L}}\sum_{k\gtrless 0}\frac{1}{\sqrt{|k|}}\left(\aop_ke^{ikx} - \adop_ke^{-ikx}\right),
\end{equation}
where $\left[\aop_k,\adop_{k'}\right]=\delta_{kk'}$ is equivalent to the chiral current $\rho_{R/L} \equiv \partial \phi_{R/L}$ satisfying the Kac--Moody algebra
\begin{equation}\label{eq:CurrentCommutator}
  [\rho_{R/L}(x),\rho_{R/L}(x')] = \mp i\delta'(x-x').
\end{equation}
The Luttinger Hamiltonian~\eqref{eq:LuttingerHam} becomes
\begin{equation}\label{eq:LuttingerHamChiralBasis}
H_\text{LL}=\frac{u}{2} \int_0^L\dd{x} \left[\rho_R^2 + \rho_L^2\right],
\end{equation}
and the chiral currents obey $[\partial_t \pm u\partial_x]\rho_{R/L} = 0$, describing decoupled right-- and left--moving density waves. 

\subsection{Shortcomings of the theory} % (fold)
\label{sub:shortcomings_of_the_theory}

% subsection shortcomings_of_the_theory (end)

The effect of the truncation of the full hydrodynamic theory Eq.~\eqref{eq:BGQHD} to the quadratic Luttinger Hamiltonian may be illustrated by considering the dynamic structure factor (DSF), defined as the (Fourier--transformed) density--density correlation function 
\begin{equation}\label{eq:DSF-def}
  S(q,\omega) = \int \dd{t}e^{i\omega t}\braket{0|\rho^{\phantom{\dagger}}_q(t)\rho^\dagger_q(0)|0}.
\end{equation}
Since the Fourier components of the chiral currents are eigenstates of the Hamiltonian \eqref{eq:LuttingerHamChiralBasis}, the DSF is destined to be a delta peak on the phonon dispersion
\begin{equation}
  S(q,\omega) = 2\pi\delta(\omega - u|q|) \braket{0|\rho_q(0)\rho^\dagger_q(0)|0}.
  \label{eq:DSF-delta} 
\end{equation}
From the Luttinger Hamiltonian Eq.~\eqref{eq:LuttingerHam} we find the ground state expectation
\begin{equation}\label{eq:densityvac}
	\braket{0|\rho^{\vphantom{\dagger}}_q\rho^\dagger_q|0}=\frac{Kq}{4\pi^2}.
\end{equation}
For a Galilean-invariant system, the weight of the delta peak is fixed by the $f$--sum rule (see, e.g., Ref.~\onlinecite{Pitaevskii:2003}), 
\begin{equation}\label{eq:f-sum-rule}
  \int \omega S(q,\omega) \dd{\omega} = n \frac{q^2}{2m},
\end{equation}
which is consistent with Eq.~\eqref{eq:densityvac} and Eq.~\eqref{eq:LuttingerParam}.

The $\delta$-function $\delta(\omega - u|q|)$ in the DSF is a consequence of the Luttinger Hamiltonian being a theory of \emph{free} phonons. The interactions between phonons that we dropped in passing from the QHD Hamiltonian Eq.~\eqref{eq:BGQHD} to Eq.~\eqref{eq:H2} are expected to a cause a broadening of this $\delta$-function, indicating a finite lifetime for the excitations. 

In more than one dimension, it is well established that the nonlinearities of QHD may be treated as a perturbation of the quadratic theory \cite{Pitaevskii:2003}. To see why such an approach is justified, recall that the QHD Hamiltonian Eq.~\eqref{eq:BGQHD} may be used to describe a Bose gas in dimensions greater than one. When expressed in terms of the creation and annihilation operators of the Fourier components, the cubic parts of the Hamiltonian have the following generic form at low momentum
\begin{equation}\label{eq:GenericCubic}
  \sum_{\bk_3=\bk_1+\bk_2\atop \{k_1,k_2,k_3\}\ll mu}\sqrt{|\bk_1||\bk_2||\bk_3|}\left[\adop_{\bk_3}\aop_{\bk_1}\aop_{\bk_2}+\adop_{\bk_1}\adop_{\bk_2}\aop_{\bk_3}\right].
\end{equation}
The use of perturbation theory is justified in the calculation of the lifetime of low energy phonons on account of the $\sqrt{|\bk|}$ factors in Eq.~\eqref{eq:GenericCubic}, which lead to vanishing interactions at long wavelengths. In superfluids the decay process due to interaction terms of this form is known as \emph{Beliaev damping}.

The kinematics of this decay process is determined by conservation of momentum and energy
\begin{equation}
  \begin{split}
    \Omega_{\bk_3} &= \Omega_{\bk_1}+\Omega_{\bk_2}\\
    \bk_3 &= \bk_1 + \bk_2.
  \end{split}  
  \label{eq:2Cons} 
\end{equation}
In classical wave physics these are sometimes known as \emph{resonance conditions}, as they determine resonant energy transfer between modes. 

Suppose that the phonon dispersion relation has the low momentum form (c.f. Eq.~\eqref{eq:BogLin})
\begin{equation}
  \Omega_\bk = u|\bk| - \beta|\bk|^3 +\ldots.
\end{equation}
For the conditions Eq.~\eqref{eq:2Cons} to be satisfied it is necessary that $\beta <0$, and that $\bk_1$ and $\bk_2$ have non-zero angle with respect to each other. Thus we see that in one dimension phonon dispersion of either sign prohibits Beliaev decay and the combined effect of dispersion and nonlinearity is necessarily more complicated.

Perhaps we can ignore the phonon dispersion? As we have seen, this means passing from the quadratic Hamiltonian $H_\text{Q}$ to the Luttinger Hamiltonian $H_\text{LL}$ in Eq.~\eqref{eq:LuttingerHam}. It is natural to consider perturbing $H_\text{LL}$ by the cubic Hamiltonian Eq.~\eqref{eq:GenericCubic} specialized to one dimension. Here the interactions come in two types: those between phonons of the same chirality
\begin{equation}\label{eq:ChiralCubicHam}
  H_{R/L} = \frac{g_{R/L}}{3!} \int_0^L \dd{x} (\partial \phi_{R/L})^3
\end{equation}
and of different chiralities
\begin{equation}\label{eq:NonChiralCubicHam}
  H_{RL} = \frac{g_{RL}}{2!}\int_0^L\dd{x} (\partial \phi_R)^2\partial\phi_L - (\partial\phi_L)^2\partial\phi_R.
\end{equation}
Such terms arise, for example, from the first term of Eq.~\eqref{eq:BGQHD}, originating from the boson kinetic energy, in which case $g_R/3! = 2\cdot g_{RL}/2! = -\frac{1}{4m}\sqrt{\frac{\pi}{2K}} = -\frac{u}{4n} \frac{1}{\sqrt{2\pi K}}$ (using the result Eq.~\eqref{eq:LuttingerParam} for the Luttinger parameter in a Galilean invariant system).

As was explicitly noted in Ref.~\onlinecite{Samokhin:1998}, perturbing $H_\text{LL}$ by the interaction between phonons of the same chirality (e.g. Eq.~\eqref{eq:ChiralCubicHam}) is problematic. To appreciate the nature of the difficulty, consider $H_R$ in the interaction picture with respect to the quadratic Hamiltonian $H_2$. In terms of the mode expansion Eq.~\eqref{eq:ChiralCurrentModeExpansion}, $H_R$ has the form
\begin{equation}
  \begin{split}
    H_R(t) \propto \sum_{k_3=k_1+k_2\atop \{k_1,k_2,k_3\}> 0}\sqrt{k_1k_2k_3}\left[\adop_{k_3}\aop_{k_1}\aop_{k_2} e^{i \left(\Omega_{k_3}-\Omega_{k_1}-\Omega_{k_2}\right)t }\right. \\
    \left.\phantom{\adop_k} + \text{h.c.}\right].
  \end{split}
\end{equation}
It is then clear that for purely linear dispersion $\Omega_k = uk$ the time dependence is absent from the Hamiltonian, and there is no sense in which $H_R$ can be regarded as a small perturbation. 

The same point can be made more compactly by noting that the chiral momentum operator
\begin{equation}\label{eq:MomentumOp}
  P = \sum_{k>0} k\adop_k\aop_k = \frac{1}{2}\int \left(\partial\phi_R\right)^2 
\end{equation}
is proportional to the chiral Luttinger Hamiltonian, $H_{\chi LL} \equiv uP$ (c.f. Eq.~\eqref{eq:LuttingerHamChiralBasis}), and a Galilean boost to the frame moving with the phonons sends $H_{\chi LL} \to H_{\chi LL} - uP = 0$, leaving only the interaction term $H_R$. 

In conclusion, there is no meaningful perturbation theory starting from the Luttinger Hamiltonian, and adding dispersion appears to forbid phonon decay on kinematic grounds. We emphasize that these are difficulties in the \emph{treatment} of the anharmonic phonon Hamiltonian: there is nothing wrong with the Hamiltonian itself.

% chiral interaction~\eqref{eq:ChiralCubicHam}. Therefore, there is no way to calculate a self--energy in perturbation theory for resonant phonons, whose imaginary part would lead to a broadening of the delta peak into a Lorentzian, as happens in higher dimensions.

\begin{figure}[t]
  \centering
  % \showthe\columnwidth
  \includegraphics[width=\columnwidth]{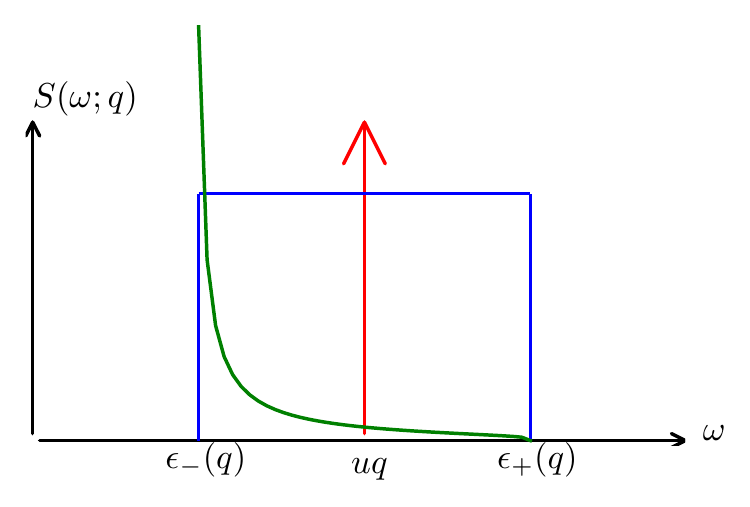}
  \caption{Results for the dynamical structure factor, Eqn.~\eqref{eq:DSF-def}, for three models: the Luttinger Liquid (red delta--peak, see Eqn.~\eqref{eq:DSF-delta}), free fermions (blue top--hat, see Eqn.~\eqref{eq:DSF-FreeFermions}), and the Calogero--Sutherland model\cite{Pustilnik:2006a} (green curve).}
  \label{fig:DSF-Predictions}
\end{figure}

As a guide, it's useful to compare the result  Eq.~\eqref{eq:DSF-delta} for the Luttinger Hamiltonian with the result for free fermions with quadratic dispersion. Fixing $q>0$, applying the density operator $\rho_q^\dagger = \sum_k\adop_{k+q}\aop_k$ to the Fermi sea creates a superposition of eigenstates of energy $\epsilon_{k,q} = [(k+q)^2-k^2]/2m=q^2/2m + kq/m$, for $k$ in the window $[k_F-q, k_F]$, where $k_F$ is the Fermi wavevector. It follows that
\begin{equation}
  \braket{\rho^{\phantom{\dagger}}_q(t)\rho^\dagger_q(0)} = \sum_{k\in [k_F-q,k_F]} e^{-iE_kt},
\end{equation}
and the structure factor has the form of a top hat (see Fig.~\ref{fig:DSF-Predictions}). 
\begin{equation}\label{eq:DSF-FreeFermions}
  S(q,\omega)=\frac{m}{q}\Theta(\omega-\epsilon_-)\Theta(\epsilon_+-\omega),
\end{equation}
where $\epsilon_\pm(q) =   k_F q/m \pm q^2/2m$. Eq.~\eqref{eq:DSF-FreeFermions} may be checked to satisfy Eqn.~\eqref{eq:f-sum-rule}.
From a distance, the top hat looks like the Luttinger liquid delta peak. However, LL theory alone is unable to resolve even this simple ``fine structure''. 

Bosonizing free fermions with cubic dispersion leads to a Luttinger Hamiltonian with chiral cubic interaction $H_{R/L}$ between phonons (Ref.~\onlinecite{Haldane:1981}, see also Appendix.~\ref{Appendix:Refermionization}). Thus we can infer that inclusion of this term in the phonon theory is responsible for passing from the $\delta$-function to the top hat. Although this a consequence of interactions between phonons, we have seen that it is difficult to explain within a simple picture of phonon decay. This is reflected by the unusual lineshape, quite different from the Lorentzian that the decay picture based on a perturbative calculation of the self-energy would suggest. By contrast, a direct calculation of the structure factor in Ref.~\onlinecite{Pereira:2007} shows that the free fermion result is reproduced to order $g_{R/L}^4$, though there is no sense in which the expansion can be truncated at this point.

In interacting models, the DSF and other correlation functions have additional singular features to be described below, and in recent years attention has focused on the development of a theory beyond the Luttinger Hamiltonian capable of describing these features \cite{Imambekov:2012}.

%, much less the power law singularities present in the exact solution for the Calogero--Sutherland model \cite{Pustilnik:2006a, Minahan:1994a, Haldane:1994, Ha:1994, Lesage:1995}, which describes particles of mass $m$ interacting via a long--range potential $U(x,y) = \frac{\lambda(\lambda-1)}{m(x-y)^2}$ \cite{Minahan:1994a, Polychronakos:1995, Awata:1995},

We are now in a position to pose the question that we address in this work: is it possible to obtain the fine structure of the DSF from an analysis of the QHD theory? 
%
%It follows from the definition Eq.~\eqref{eq:DSF-def} that to obtain anything other than a delta peak requires that we account for interactions between phonons.
%
\begin{comment}

\begin{equation}\label{eq:Self-Energy}
  \Sigma(\epsilon(q), q) = \frac{g^2}{2L}\sum_{k}\frac{kq(q-k)}{\epsilon(q)-\epsilon(k)-\epsilon(q-k)}.
\end{equation}

\end{comment}
%
% On the other hand, interactions between different chiral sectors are non--resonant, giving a finite self--energy. However (see below) the quasiparticle residue is vanishing provided the left--movers are gapless (c.f. the edge waves in the quantum Hall effect), or $g_{RL}$ is otherwise absent (as occurs in the Calogero--Sutherland model), so a simple perturbative calculation in $g_{RL}k/c$ now falls over the second hurdle.  
%
%It turns out that the interaction of dispersive phonons with both co- and counter-propagating modes, though non-resonant, leads to a singular renormalization of the phonon quasiparticle weight. 
%
Before turning to our approach we first describe how these difficulties were handled in earlier work.

\subsection{Earlier work}

Due to the resonances plaguing the bosonic perturbation theory about the Luttinger Hamiltonian, Rozhkov \cite{Rozhkov:2005} was led to a basis of fermionic quasiparticles. As we discussed above, the QHD Hamiltonian $H_{\text{LL}} + H_R + H_L$ can be described \emph{exactly} in terms of \emph{free} fermions with quadratic dispersion, and this proves to be a convenient starting point for the analysis of the more general case (including dispersion and $H_{RL}$), which will correspond to \emph{interacting} fermions.

Ref.~\onlinecite{Pustilnik:2006} made the following key observation with respect to this situation. The free fermion result Eq.~\eqref{eq:DSF-FreeFermions} for $S(q,\omega)$ is characterized by the presence of upper and lower thresholds $\epsilon_\pm$(q) that delimit the range of allowed energies of a particle--hole pair of momentum $q$. The upper threshold corresponds to the creation of a particle--hole pair with the particle at momentum $k_F+q$ and the hole at $k_F$ i.e. the Fermi surface, while the lower threshold has the particle at $k_F$ and the hole at $k_F-q$. If we are concerned with the behaviour of the DSF at energies that differ from the upper (lower) threshold by an amount $\ll k_F q/m$, we may regard the particle (hole) as being a distinct excitation, interacting with a Luttinger Liquid.

%Bosonizing free fermions with finite band curvature leads to a Luttinger Hamiltonian with chiral cubic interaction $H_{R/L}$ between phonons (\cite{Haldane:1981}, see also Appendix.~\ref{Appendix:Refermionization} \todo{???}). Running this reasoning backwards, the structure factor for a Hamiltonian with a chiral cubic interaction must be the free fermion ``top hat'' of Eq.~\eqref{eq:DSF-FreeFermions} (see Fig.~\ref{fig:DSF-Predictions}). 

%This result is valid when (1) we can neglect the dispersion of the phonons (the opposite regime to the semiclassical limit that we will address), and (2) the Luttinger $K$ is close to one, so that the residual cubic interaction may be treated perturbatively.
%\todo{It's exact, no?}
%A by constructing a unitary transformation mapping interacting electrons onto weakly-interacting fermions with band curvature, where perturbation theory is well-defined provided the microscopic interaction (which translates to $K \sim 1$) is sufficiently small. 
The universality of the physics near the threshold $\epsilon_{-}(q)$ \ was encapsulated by the construction of a minimal Hamiltonian \cite{Pustilnik:2006,Imambekov:2012}, known as the ``mobile impurity model''. 

We will discuss only the hole case for brevity. The important degrees of freedom are isolated by writing the microscopic fermion field $\pdop(x)$ in terms of excitations near the right and left Fermi points, and a ``deep hole'', or mobile impurity, created by $\dop(x)$ with momentum $\sim k_F-q$, 
\begin{equation}\label{eq:FermionProjection}
  \pop(x) \to e^{ik_Fx}\pop_R(x) + e^{-ik_Fx}\pop_L(x) + e^{i(k_F-q)x} \dop(x).
\end{equation}
$\pop_{R/L}$ are described by the Luttinger Hamiltonian, and coupled to the impurity via
\begin{equation}\label{eq:MobileImpurityHam}
  \begin{split}
    H_{\dop} &= \int \dd{x} \ddop(x)[\epsilon_{-}(q) - iv_d\partial_x]\dop(x), \\
    H_{\text{int}} &= \int \dd{x} [V_R(q)\rho_R(x) + V_L(q)\rho_L(x)]\dop\ddop.
  \end{split}
\end{equation}
Here $v_d = \epsilon'_-(q)$ is the group velocity of the hole.
Just as with the linear Luttinger Liquid, the parameters $V_{R/L}$ must be found from the microscopic model. If the microscopic interactions are small, the model~\eqref{eq:MobileImpurityHam} may be derived with explicit expressions for the coupling parameters. Otherwise, the model should be regarded as phenomenological, and the parameters found by calculating its response to shifts in the zero modes \cite{Imambekov:2012}. 
Collecting those terms from Eqn.~\eqref{eq:FermionProjection} that contribute terms peaked around $e^{-iqx}$ to the density $\pdop\pop$, we find the DSF (Eq.~\eqref{eq:DSF-def}) may be written 
\begin{equation}\label{eq:DSF-projected}
  \int \dd{t} e^{i\omega t}\braket{\ddop(x,t)\pop_R(x,t)\pdop_R(0,0)\dop(0,0)}_{H_0+H_d+H_{\text{int}}}
\end{equation}
A unitary transformation removes $H_{\text{int}}$, just as in the bosonization solution of the X--ray edge problem \cite{Schotte:1969}. Whilst in the pure X--ray edge problem the impurity is non--dynamical, the dynamics of our impurity can be accounted for simply by shifting to the frame moving with $v_d$. Applying the unitary transformation
\begin{equation}
  U_R = \exp\left(i\int \frac{V_R(q)}{u-v_d}\phi_R\dop\ddop \dd{x}\right)
\end{equation}
shifts $\rho_R \to \rho_R + V_R(q)\ddop\dop/(u-v_d)$ (the combination $u-v_d$ appears due to the shift to the impurity frame). A similar transformation for the left--movers removes the interaction term from the Hamiltonian. 

To find the action of $U_R$ on the projected electron operators $\pdop_R$ we use the bosonization formula  $\pdop_{R/L} \sim e^{i[\phi\mp\theta]}$.\cite{Giamarchi:2004} To lighten the algebra, we concentrate on the simple case $K=1$, where the $\pdop_{R/L}$ are just $e^{i\phi_{R/L}}$. After the transformation, the DSF \eqref{eq:DSF-projected} factorises into the impurity and LL correlators

\begin{comment}
In terms of the chiral fields,
%
\begin{equation}
  \begin{split}
    \pdop_R &\sim \exp \left( i\sqrt{\frac{\pi}{2}}\left[\left(\frac{1}{\sqrt{K}}+\sqrt{K}\right)\phi_R + \left(\frac{1}{\sqrt{K}}-\sqrt{K}\right)\phi_L \right]\right)\\
    \pdop_L &\sim \exp \left( i\sqrt{\frac{\pi}{2}}\left[\left(\frac{1}{\sqrt{K}}-\sqrt{K}\right)\phi_R + \left(\frac{1}{\sqrt{K}}+\sqrt{K}\right)\phi_L\right] \right)
  \end{split}
\end{equation}
\end{comment}
%
\begin{equation}
  \begin{split}
    &\int \dd{t}e^{i\omega t}\Braket{\exp \left(i\delta_R[\phi_R(x,t)-\phi_R(0,0)]\right)}_{H_0}\\
    &\times\Braket{\exp\left(i\delta_L[\phi_L(x,t)-\phi_L(0,0)]\right)}_{H_0}\\
    &\times \braket{\ddop(x,t)\dop(0,0)}_{H_d}.
  \end{split}
\end{equation}
We used the shorthand $\delta_{L/R}$ for $V_R/(u-v_d), V_L/(u+v_d)$, which are interpreted as momentum dependent phase shifts for the impurity scattering on the Luttinger liquid. Using the results for correlation functions of vertex operators in the Luttinger liquid\cite{Giamarchi:2004}, and then taking a Fourier transform, produces power law behaviour for the structure factor
\begin{equation}\label{eq:DSF-mobile-impurity-result}
  S(q,\omega) \propto  [\omega - \epsilon_-(q)]^{-\mu(q)}\Theta(\omega - \epsilon_-(q)), \hspace{5mm} \omega \sim \epsilon_{-}(q),
\end{equation}
where
\begin{equation}\label{eq:DSF-mobile-impurity-exponents}
  \mu(q) = 1-\frac{\delta_L^2}{2\pi}-\frac{\delta_R^2}{2\pi}. 
\end{equation}
Eq.~\eqref{eq:DSF-mobile-impurity-result} predicts a power law behaviour of $S(q,\omega)$ near the lower threshold, with $q$-dependent exponent. The physical picture is that the hole acts a an impurity with momentum close to $k_F-q$, causing a shake up of the Luttinger liquid thanks to Anderson's orthogonality catastrophe \cite{Anderson:1967}: the ground state of the Luttinger liquid in the presence of the impurity is orthogonal (in the large--system limit) to the ground state in its absence. This softens the delta--function propagator of the impurity into a power law singularity.

As we have already mentioned, the mobile impurity model should be regarded as a phenomenological description of the threshold singularities outside of the weak interaction regime. Useful checks on the correctness of this picture are, however, available from exactly--solvable models. For instance, the Calogero--Sutherland model\cite{Minahan:1994a, Polychronakos:1995, Awata:1995}, which describes particles of mass $m$ interacting via a long--range potential $U(x,y) = \frac{\lambda(\lambda-1)}{m(x-y)^2}$, has exponents $\mu_{+} = 1-\lambda$, $\mu_- = 1-1/\lambda$.\cite{Pustilnik:2006a}

Last but by no means least, broadening of the dynamic structure factor beyond any effects attributable to the finite temperature has recently been observed in experiments on a 1D gas of cold $\mathrm{^{87}Rb}$ atoms.\cite{Inguscio:2015}

\subsection{Scope of this paper} % (fold)

% subsection scope_of_this_paper (end)

It is the purpose of this paper to calculate the (zero temperature) structure factor directly from the hydrodynamic Hamiltonian containing the leading nonlinear corrections to the dispersion and interactions of the Luttinger Hamiltonian, Eq.~\eqref{eq:LuttingerHam}. 

As we have explained, it is necessary to include both interactions between phonons described by~\eqref{eq:ChiralCubicHam}, \eqref{eq:NonChiralCubicHam}, and dispersion $\Omega(q) = uq + \epsilon(q)$, which we model by the expansion
\begin{equation}\label{eq:PhononDispersion}
  \epsilon(q)= - \alpha q^2 - \beta q^3+\ldots.
\end{equation}
For the Bogoliubov dispersion Eq.~\eqref{eq:BogFreq} $\alpha=0$, $\beta<0$. An example of a model containing a $q^2$ dispersion is provided by the aforementioned Calogero--Sutherland model.  Bosonization of the $1/r^2$ interaction gives rise to a term 
\begin{equation}
  %\frac{\lambda(\lambda-1)}{2m}\sum_{q>0}q^2\phi_q\phi_{-q}.
  %\todo{Isn't it $q^3$?}
  \frac{\lambda(\lambda-1)}{2m \pi^2}\int \dd{x} \dd{y} \frac{\partial\phi(x)\partial\phi(y)}{(x-y)^2}.
  \label{eq:CSBosonized} 
\end{equation}
For this model, $\alpha = \frac{(\lambda - 1)}{2m}, g_L=g_R = \frac{\sqrt{2\pi\lambda}}{m}$, and coupling between right and left movers is absent $g_{RL}=0$ (see, e.g. Ref.~\onlinecite{stone2008quantum} for a clear discussion of this point). 

The term Eq.~\eqref{eq:CSBosonized} may be rewritten
\begin{equation}
  \frac{\lambda(\lambda-1)}{2m \pi}\int \dd{x}  \partial\phi\partial^2\phi_H,
\end{equation}
where the subscript indicates the Hilbert transform,
\begin{equation}\label{eq:HilbertDef}
  f_H(x) \equiv \frac{\PV}{\pi}\int_{-\infty}^\infty \frac{f(y)}{y-x}\dd{y}.
\end{equation} 
To recap, the low--energy phonons are governed by the Hamiltonian consisting of the dispersion \eqref{eq:PhononDispersion}, chiral interactions \eqref{eq:ChiralCubicHam}, and nonchiral interactions \eqref{eq:NonChiralCubicHam}. In a theory of classical 1D waves, an initial disturbance will spatially separate into left--moving and right--moving profiles. For this reason, classical nonlinear wave equations are usually written for waves moving in one direction in a frame moving at the speed of $k\to 0$ waves and with only the dispersive terms retained. Similarly, we would like to work with chiral excitations only, but we must be careful to first remove the interaction $H_{RL}$ (see Eq.~\eqref{eq:NonChiralCubicHam}) by a Schrieffer--Wolff transformation (discussed in Section~\ref{sec:ChiralMap}). After moving to a frame with velocity equal to the speed of sound, we arrive at the chiral Hamiltonian for right moving excitations 
\begin{equation}\label{eq:PhononHam}
  H_\chi = \int \dd{x} \left[ \frac{\alpha}{2}\partial\phi\partial^2\phi_H - \frac{\beta}{2}(\partial^2\phi)^2 + \frac{g}{3!}(\partial\phi)^3\right],
\end{equation}
where we have dropped the subscript $R$ to lighten the notation, as well as the term $(u/2) \left(\partial\phi\right)^2$ describing the linear dispersion which, as we saw, can be removed by passing to frame with velocity $u$.

The Hamiltonian Eq.~\eqref{eq:PhononHam} is the central object of study in this paper. A striking feature of chiral Hamiltonians~\cite{Sakita:1985} of this form is that, within a sector of fixed momentum $P$, the spectrum is bounded from below \emph{and} above. To understand why this is so, first note that for $g=0$ we have a theory of free chiral bosons with dispersion given by Eq.~\eqref{eq:PhononDispersion}
\begin{equation}
H_{\chi,g=0} = \sum_k \epsilon(k)\adop_k\aop_k. 
\label{eq:PhononHamZeroInteraction} 
\end{equation}
At fixed momentum $P=\sum k \adop_k\aop_k=q$, the spectrum of $H_{\chi,g=0}$ lies in the range $[\epsilon(q),0)$. $\epsilon(q)$ corresponds to a single phonon of momentum $q$. Conversely, the energy can be made arbitrarily close to zero by partitioning $q$ among many low momentum modes. Thus the quadratic part of the Hamiltonian is bounded. The same goes for the cubic interaction term: we know that this term describes free chiral fermions with quadratic dispersion, which as we have seen have a bounded spectrum at fixed momentum. Eq.~\eqref{eq:PhononHam}, being the sum of two operators that are bounded above and below, then shares this property.

\begin{figure}[t]
  \centering
  % \showthe\columnwidth
  \includegraphics[width=\columnwidth]{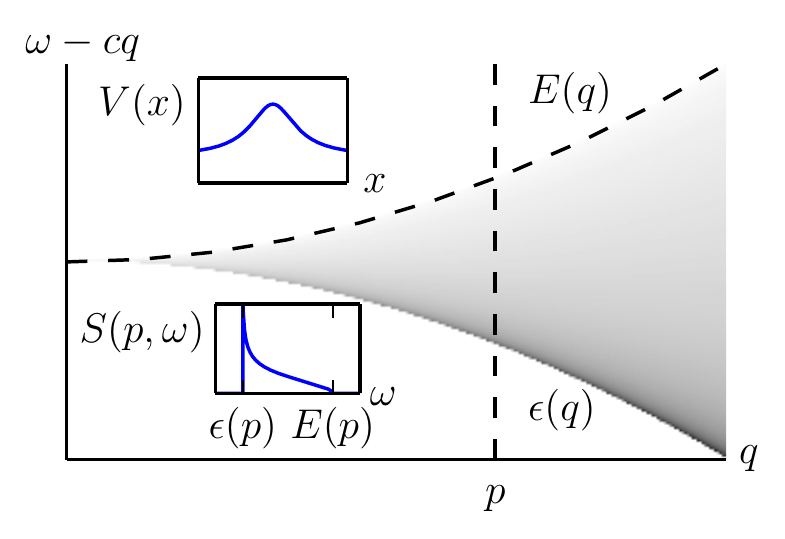}
  \caption{For the chiral Hamiltonian, the dynamical structure factor (greyscale) has power--law singularities given by Eqn.~\eqref{eq:DSF-chiral-result} in the vicinity of the thresholds given by the phonon dispersion $\epsilon(q)$ and soliton dispersion $E(q)$ (Eqns.~\eqref{eq:PhononDispersion}, \eqref{eq:SolitonDispersion}).}
  \label{fig:ChiralStrucFac}
\end{figure}

The DSF of Eq.~\eqref{eq:PhononHam} will be finitely supported between two thresholds. Central to our approach is the identification of a parameter that permits a semiclassical interpretation of these thresholds. One is a phonon dispersion relation, appropriately renormalized by interactions, as we will see in more detail in Section~\ref{sec:Phonon}. The other is connected to \emph{soliton} solutions of the equation of motion of Eq.~\eqref{eq:PhononHam}. Writing $v\equiv g\partial \phi$, this has the form
\begin{equation}\label{eq:BO-KdV-eqn}
  0 = \partial_tv + v\partial_x v + \alpha \partial_x^2v_H + \beta \partial_x^3v. 
\end{equation}
When $\alpha = 0$, this is the Korteweg--de Vries equation of fluid dynamics. The case $\beta = 0$ is known as the Benjamin--Ono equation \cite{Benjamin:1967,Ono:1975}. Both equations are integrable \cite{Fokas:1983,Coifman:1990,Kaup:1998} (although the mixed case with both $\alpha, \beta \neq 0$ is not) and have a family of soliton solutions parameterized by a soliton speed $v_S$ (which follows the sign of $\alpha$, $\beta$ respectively)
\begin{equation}\label{eq:SolitonSolution}
  V(x-v_St) = \begin{cases}
    3v_S \sech^2\left(\frac{1}{2}\sqrt{\frac{v_S}{\beta}}x\right) & \text{(KdV)},\\
    \frac{4\alpha^2v_S}{v_S^2x^2+\alpha^2} & \text{(BO)}.
  \end{cases}
\end{equation}
Substituting the soliton solutions into Eqns.~\eqref{eq:BO-KdV-eqn} and~\eqref{eq:MomentumOp}, we find the dispersion relations
\begin{equation}\label{eq:SolitonDispersion}
  E(q) = \begin{cases}
    \frac{3}{5}\left(\frac{g^2}{12\sqrt{\beta}}\right)^{2/3}P^{5/3} & \text{(KdV)},\\
    \frac{g^2}{8\pi \alpha} P^2 & \text{(BO)}.
  \end{cases}
\end{equation}
Note that these bend in the opposite sense to the phonon dispersion, as shown in Fig.~\ref{fig:ChiralStrucFac}. 

%In the chiral model, these dispersions correspond to the maximum and minimum energy states at a given momentum. To see this, one has to calculate matrix elements of the Hamiltonian among the set of states with the correct momentum. \todo{I think this needs to be expanded (or cut)}

The identification of a threshold in the DSF with a soliton dispersion relation appears in the context of the 1D Bose gas (Lieb--Liniger model) in Ref.~\onlinecite{Khodas:2008}. Note that in a nonchiral model only a lower threshold exists, as one can make arbitrarily energetic states at fixed momentum by creating both left and right movers. Nevertheless, we will see that we can understand the DSF in the nonchiral case by carefully decoupling the two chiralities.

We now state the main results of this paper, the behaviour of the structure factor in the vicinity of the thresholds $\epsilon(q), E(q)$, in the semiclassical regime where dispersion dominates the interaction. This requires $g_Rq/\epsilon'(q), g_{RL}q/2u \ll 1$. In the opposite limit, at least for $g_{RL}$ vanishing, refermionization gives a top hat for the structure factor \cite{Rozhkov:2005}.

In the chiral problem, the structure factor has support only within the window of frequencies $[\epsilon(q),E(q)]$, and has power law behaviour at the thresholds $\epsilon(q), E(q)$ given by
\begin{equation}\label{eq:DSF-chiral-result}
  S(q,\omega) \propto \begin{cases}
    [\omega - \epsilon(q)]^{\mu_R(q)-1} & \omega \gtrsim \epsilon(q)\\
    [E(q)-\omega]^{\Delta^2/2\pi -1} & \omega \lesssim E(q),
  \end{cases}
\end{equation}
(see Fig.~\ref{fig:ChiralStrucFac}). The exponents are determined by $\mu_R(q) = \frac{1}{2\pi}\left(g_Rq/\epsilon'(q)\right)^2$, and the soliton weight $\Delta(q) = \int \dd{x} V$ respectively. Strictly speaking, the thresholds should contain the renormalized dispersion parameters. However, these corrections may be calculated using conventional perturbation theory, so we do not discuss them in this paper.

\begin{figure}[t]
  \centering
  % \showthe\columnwidth
  \includegraphics[width=\columnwidth]{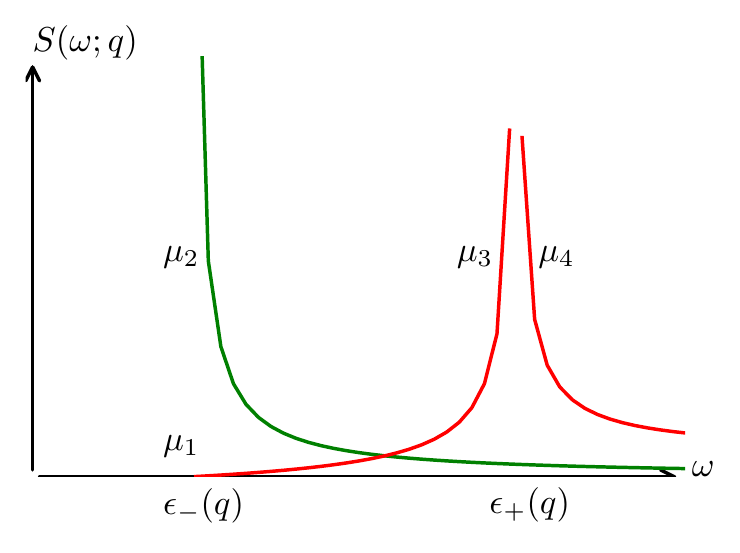}
  \caption{For the chiral Hamiltonian Eq.~\eqref{eq:PhononHam}, the structure factor takes the form of the red curve if the phonons define the upper threshold, and the green curve if the soliton defines the upper threshold. See Eqn.~\eqref{eq:NonchiralExponents} for the values of the exponents $\mu$.}
  \label{fig:NonChiralStrucFac}
\end{figure}

If there are interactions with gapless left--movers, the structure factor is also non--zero above the upper threshold, since excitations of arbitrarily high energy at fixed momenta may be produced. There are two cases, as shown in Fig.~\ref{fig:NonChiralStrucFac}, depending on whether the phonon or soliton defines the lower threshold. The values of the exponents are
\begin{equation}\label{eq:NonchiralExponents}
  \begin{split}
    \mu_1 &= \frac{1}{2\pi}\Delta^2(q) + \frac{1}{2\pi}\left(\frac{g_{RL}q}{2u}\right)^2 - 1, \\
    \mu_2 &= \frac{1}{2\pi}\left(\frac{g_Rq}{\epsilon'(q)}\right)^2 + \frac{1}{2\pi}\left(\frac{g_{RL}q}{2u}\right)^2-1,\\
    \mu_3 &= \frac{1}{2\pi}\left(\frac{g_{R}q}{\epsilon'(q)}\right)^2-1,\\
    \mu_4 &= \frac{1}{2\pi}\left(\frac{g_{RL}q}{2u}\right)^2-1.
  \end{split}
\end{equation}

\begin{comment}

If the phonons define the lower threshold, the structure factor is analytic at the upper threshold, and diverges as
%
\begin{equation}\label{eq:Nonchiral-phonon-lower-power}
  [\omega-\epsilon(q)]^{\mu_R+\mu_{RL} - 1}\Theta(\omega -\epsilon(q)), \hspace{10mm} \omega \sim \epsilon(q)
\end{equation}
%
at the lower threshold, and $\mu_{RL}(q) = \frac{1}{2\pi}\left(g_{RL}q/2c\right)^2$ . If the phonons define the upper threshold, the structure factor has a two--sided power law at the upper threshold
%
\begin{equation}\label{eq:Nonchiral-phonon-upper-powers}
  \begin{cases}
    [\omega-\epsilon(q)]^{\mu_{RL}-1}, & \omega \gtrsim \epsilon(q)\\
    [\epsilon(q)-\omega]^{\mu_R-1,} & \omega \lesssim \epsilon(q),
  \end{cases}
\end{equation}
%
and a one--sided power law at the lower threshold given by the soliton
%
\begin{equation}\label{eq:Nonchiral-soliton-lower-power}
  [\omega-E(q)]^{\Delta^2/2\pi + \mu_{RL} -1}\Theta(\omega-E(q)), \hspace{10mm} \omega \sim E(q).
\end{equation}
%
\end{comment}

\subsection{Outline} % (fold)
\label{sub:outline}

To conclude this introduction, we describe in outline the remainder of the paper. In Section~\ref{sec:ChiralMap} we show how to decouple the interaction between left-- and right--movers via a unitary transformation, which expresses correlation functions as a convolution of a correlator in a (nonlinear) chiral Hamiltonian with a power law originating from the chiral decoupling. Correlation functions in the chiral Hamiltonian have power law singularities near the phonon threshold, which are calculated in Section~\ref{sec:Phonon}, and soliton threshold, which are calculated in Section~\ref{sec:Soliton}.  The calculations for a purely chiral theory with quadratic phonon dispersion are discussed in a previous paper by the authors\cite{Price:2014}.

% subsection outline (end)

\section{Mapping to a chiral problem}
\label{sec:ChiralMap} 

The Luttinger Hamiltonian Eq.~\eqref{eq:LuttingerHamChiralBasis} describes free left and right moving phonons. The two branches are coupled together by the term $H_{RL}$ (see Eq.~\eqref{eq:NonChiralCubicHam}). In this section we will show how to approximately decouple the two branches in the presence of this interaction. Despite being non--resonant, it has an important influence on the threshold singularities of the DSF, because the Anderson orthogonality catastrophe causes a vanishing of the quasiparticle residue in the thermodynamic limit.

\begin{figure}[t]
  \centering
   \centering
   \begin{subfigure}[b]{0.4\columnwidth}
     \includegraphics[width=\textwidth]{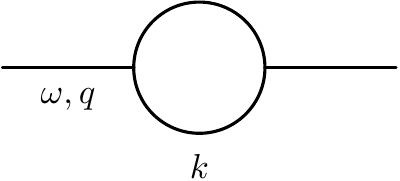}
     \caption{Self--energy bubble\label{fig:FullOneLoop}}
   \end{subfigure}
   \begin{subfigure}[b]{0.4\columnwidth}
     \includegraphics[width=\textwidth]{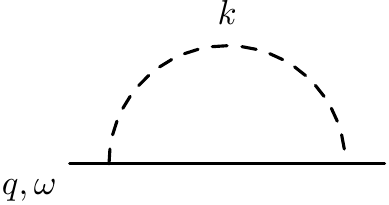}
     \caption{Leading logarithm approximation\label{fig:SoftOneLoop}}
   \end{subfigure}
  \caption{The contribution to the phonon self--energy at one loop, including all loop momenta \ref{fig:FullOneLoop}, and its leading logarithmic divergence \ref{fig:SoftOneLoop}, where the dashed propagator has momenta $k\ll \Lambda$. \label{fig:Self-energy}}
\end{figure}

We work with the Hamiltonian $H_\text{LL} + H_{RL}$, ignoring dispersion for the moment, which does not effect the singularity. Second order perturbation theory in $H_{RL}$ gives a self--energy for the right moving phonon corresponding to the diagram in Fig.~\ref{fig:FullOneLoop} (the loop containing a left and a right mover),
\begin{equation}
  \Sigma(\omega, q>0) = \frac{g_{RL}^2}{2L}\sum_{k<0} \frac{|kq(q-k)|}{\omega - u|q-k|-u|k|}.
\end{equation}
Due to the $\omega$ dependence, the resulting Green function
\begin{equation}
  \frac{1}{\omega - \epsilon(q) -\Sigma(\omega,q)} 
\end{equation}
no longer has unit residue at its pole, but is down by a factor $Z$ (the \emph{quasiparticle residue}), where 
\begin{equation}
  Z^{-1} = 1 - \left.\pbyp{\Sigma}{\omega} \right|_{\omega = \epsilon(q)}.
\end{equation}
The self--energy derivative on shell is
\begin{equation}
  \begin{split}
    \left.\pbyp{\Sigma}{\omega}\right|_{\omega = uq} &= -\frac{g_{RL}^2}{2L}\sum_{k>0}\frac{kq(k+q)}{(2uk)^2}\\
    &=-\frac{g_{RL}^2q}{8u^2} \frac{1}{L}\sum_{k>0} \left(\frac{q}{k} + 1\right).
%    &\sim -\frac{g_{RL}^2q^2}{16\pi c^2}\int_{L^{-1}}^1\frac{\dd{k}}{k}\\
%    &= \frac{g_{RL}^2q^2}{16\pi c^2}\log L\Lambda_{RL}.
  \end{split}
\end{equation}
We are interested in the limit $qL \to \infty$, where the derivative is log divergent. Retaining only this divergence in $qL$ corresponds to approximating diagram in Fig.~\ref{fig:FullOneLoop} with that in Fig.~\ref{fig:SoftOneLoop}, and gives
\begin{equation}
  \begin{split}
    \left.  \pbyp{\Sigma}{\omega}\right|_{\omega = uq} &\sim -\frac{g_{RL}^2q^2}{16\pi u^2}\int_{2\pi/L}^q \frac{\dd{k}}{k} \\
    &\sim -\frac{g_{RL}^2q^2}{16\pi u^2}\log qL.
  \end{split}
\end{equation}
From this result we find the quasiparticle residue is
\begin{equation}
  Z = \frac{1}{1 + \frac{g_{RL}^2q^2}{16\pi u^2}\log qL},
\end{equation}
and vanishes in the IR. 
Consequently, even if the dimensionless interaction strength $g_{RL}q/u$ is small, the phonons are not good quasiparticles. Such a result holds for interactions between any non--resonant gapless modes in 1D (a fact that will be of importance when we deal with the chiral case). The vanishing of the quasiparticle residue indicates the Anderson orthogonality catastrophe \cite{Anderson:1967}. Here, the right--moving phonon plays the role of the impurity, since in the leading logarithm approximation we consider only those interactions with the soft left--movers (see Fig.~\ref{fig:SoftOneLoop}), where the momenta of the right--mover is changed very little.

To proceed, we make a unitary transformation on our operators and states generated by $A_{RL}$ that removes the interaction to leading order in $g_{RL}q/\epsilon'(q)$. Anticipating that $A_{RL}$ will be of leading order in $g_{RL}q/\epsilon'(q)$ (justified \emph{a posteriori}), and requiring that the transformed Hamiltonian contains no interactions coupling left and right movers at leading order, we find that $A_{RL}$ must satisfy
\begin{equation}
  \begin{split}
    &\phantom{=}e^{A_{RL}}\left(H_\text{LL} + H_{RL} + H_{R}\right)e^{-A_{RL}}\\
    &= H_\text{LL} + \underbrace{[H_\text{LL},A_{RL}] + H_{RL}}_{\mbeq 0} + H_R + O\left([g_{RL}q/\epsilon']^2\right).
  \end{split}
\end{equation}
Using the form of Eqn.~\eqref{eq:NonChiralCubicHam}, the appropriate generator is
\begin{equation}
  \begin{split}
    A_{RL} &= \frac{g_{RL}}{2u}\frac{1}{\sqrt{L}}\sum_{k,q>0}\sqrt{\frac{q(k+q)}{k}}\left( \adop_q\aop_{-k}\aop_{q+k} \right.\\
    &\phantom{=\sum}\left. + \adop_{-q}\aop_k\aop_{-k-q} - \hc \right).
\end{split}
\end{equation}
This has singular matrix elements coupling states with a single right--mover momentum $q$, and a slow left--mover of momentum $-k$ and right--mover $q+k$. Acting on a right--mover, and summing over only those singular interactions where $k\ll q$, the commutator simplifies to
\begin{equation}
  [A_{RL},\adop_q] = \frac{g_{RL}q}{2u}\frac{1}{\sqrt{L}}\sum_{k>0}\frac{1}{\sqrt{k}}\left(\aop_{-k}\adop_{q-k} - \adop_{-k}\adop_{q+k}\right)
\end{equation}
To calculate the action of $A_{RL}$ on the right--moving chiral boson, we have to split it into its positive and negative Fourier modes
\begin{equation}
  \begin{split}
    \phi_R^- &= i\frac{1}{\sqrt{L}}\sum_{q>0}\frac{1}{\sqrt{q}}\adop_qe^{-iqx},\\
    \phi_R^+ &= -i\frac{1}{\sqrt{L}}\sum_{q>0}\frac{1}{\sqrt{q}}\aop_qe^{iqx},
  \end{split}
\end{equation}
from which follows 
\begin{equation}
  [A_{RL},\phi_R^-(x)] = i\frac{1}{\sqrt{L}}\sum_{q>0} \left(i\frac{g_{RL}u}{2c}\phi_L(x)\right)\frac{1}{\sqrt{q}}\adop_qe^{-iqx}.
\end{equation}
Neglecting the (non--singular) commutator of $A$ and $\phi_L(x)$, we can sum the series to calculate the rotated right mover $e^{A_{RL}}\phi_R^-(x)e^{-A_{RL}}$, which takes the pleasing form
\begin{equation}
  i\frac{1}{\sqrt{L}}\sum_{q>0}\exp \left(i\frac{g_{RL}q}{2u}\phi_L(x)\right)\frac{1}{\sqrt{q}}\adop_qe^{-iqx}.
\end{equation}
Similarly, the positive modes are rotated to
\begin{equation}
  -i\frac{1}{\sqrt{L}}\sum_{q>0}\exp \left(-i\frac{g_{RL}q}{2u}\phi_L(x)\right)\frac{1}{\sqrt{q}}\aop_qe^{iqx}.
\end{equation}
The free vacuum $\ket{0}$ defined by $\aop_k\ket{0} = 0$ is invariant under $A_{RL}$ because it is annihilated by the cubic interaction Hamiltonian. Therefore, to calculate the structure factor, we substitute our rotated operators into the correlator $\braket{0|\rho^+(x,t)\rho^-(0,0)|0}_{H_\text{LL}+H_\chi+H_{RL}}$ to find it is 
\begin{equation}\label{eq:ChiralDecoupledDensityCorrelator}
  \begin{split}
    &\sum_{q>0} \Braket{\exp\left(-i\frac{g_{RL}q}{2u}\phi_L(x,t)\right)\exp\left(i\frac{g_{RL}q}{2u}\phi_L(0,0)\right)}_{H_\text{LL}}\\
    &\phantom{\sum}\times qe^{iqx}\braket{\aop_q(t)\adop_q(0)}_{H_\text{LL}+H_\chi}.
  \end{split}
\end{equation}
The factorization into left-- and right--moving correlators occurs at leading order in $g_{RL}q/u$, where the Hamiltonian does not mix the two sectors. The left--moving correlator is $\sim(i[x+ut-i0])^{-\mu_{RL}}$, where $\mu_{RL}(q) = \frac{1}{2\pi}(g_{RL}q/2u)^2$. To show this, we use the mode expansion \eqref{eq:ChiralCurrentModeExpansion} for the left movers, and use the Baker--Campbell--Hausdorff formula to disentangle the exponentials. The $x,t$ dependence comes from the part
\begin{equation}
  \begin{split}
    &\left\langle \exp\left(-\frac{g_{RL}q}{2u\sqrt{L}}\sum_{k<0}\frac{1}{\sqrt{|k|}}\aop_ke^{ikx-iu|k|t}\right)\right.\\
        &\phantom{\langle}\left.\times\exp\left(-\frac{g_{RL}q}{2u\sqrt{L}}\sum_{k<0}\frac{1}{\sqrt{|k|}}\adop_k\right) \right\rangle\\
%    &= \exp \left(\frac{1}{2\pi}\left(\frac{g_{RL}q}{2c}\right)^2\sum_{n>0}\frac{1}{n}e^{2\pi i[ -x -ct]n/L}  \right)\\
    &= \exp \left( -\frac{1}{2\pi}\left(\frac{g_{RL}q}{2u}\right)^2\log\left[1-e^{-2\pi i(x+ut-i0)/L} \right] \right)\\
    &\approx \left(2\pi i[x+ut-i0]\right)^{-\mu_{RL}(q)}.
  \end{split}
\end{equation}
The infinitesimal $i0$ makes this quantity well--defined and ensures it is built out of negative wavevectors and positive energies. We now turn to an analysis of the right--moving correlator $q\braket{\adop_q(t)\aop_q(0)}$ near the phonon and soliton thresholds.

\section{Phonon Threshold}
\label{sec:Phonon} 

We now discuss the calculation of the chiral correlation function $\braket{\aop_q(t)\adop_q(0))}_{H_\chi}$ appearing in the full nonchiral correlator, Eqn.~\eqref{eq:ChiralDecoupledDensityCorrelator}. In the limit $gq/\epsilon'(q) \ll 1$, that is where the dispersion dominates the interaction, we can calculate perturbatively around the phonon. For dispersions growing faster than $q\log q$, this is true only at sufficiently high momentum, except in the degenerate Benjamin--Ono case where it only holds values of the parameters $g/\alpha \ll 1$.

It is now convenient to split the chiral Hamiltonian of Eqn.~\eqref{eq:PhononHam} into its quadratic dispersive term $H_2$ and cubic interaction term $H_3$. As ever, it is convenient to carry out the calculation in momentum space, where the interaction reads
\begin{equation}\label{eq:MomentumSpaceInteraction}
  H_3 = \frac{g}{2\sqrt{L}}\sum_{k_i>0} \sqrt{k_1k_2k_3}\left(\adop_{k_1}\aop_{k_2}\aop_{k_3} + \adop_{k_1}\adop_{k_2}\aop_{k_3}\right).
\end{equation}
We note that the free vacuum $\ket{0}$ remains an eigenstate at zero energy of the Hamiltonian~\eqref{eq:MomentumSpaceInteraction}.  

At second--order perturbation theory, there is a contribution to the self--energy corresponding to the diagram in Fig.~\ref{fig:Self-energy} (the loop now understood to contain only right--moving phonons) with the result
\begin{equation}\label{eq:OneLoopSelfEnergy}
  \Sigma(\omega, q) = \frac{g^2}{2L}\sum_{0<k<q}\frac{kq(q-k)}{\omega-\epsilon(k)-\epsilon(k-q)}.
\end{equation}

On shell, there is a finite correction to the dispersion. However, the quasiparticle residue,  $Z^{-1} = 1-\left.\pbyp{\Sigma}{\omega}\right|_{\omega = \epsilon(q)}$, vanishes logarithmically with system size, indicating that the true quasiparticles have zero overlap with the phonons. Again this arises from the low $k$ contributions, where the energy dominator $\epsilon(q)-\epsilon(k)-\epsilon(q-k) \approx k\epsilon'(q)$, and 
\begin{equation}
  \left.\pbyp{\Sigma}{\omega}\right|_{\omega = \epsilon(q)} \approx -\frac{g^2}{2L}\sum_{0<k<q} \frac{q^2}{k[\epsilon'(q)]^2} = -\frac{1}{2\pi }\left(\frac{gq}{\epsilon'(q)}\right)^2\log  qL.
\end{equation}
Just as in the ambichiral case, the $\adop_q$ are not good quasiparticles even if the dimensionless interaction strength $gq/\epsilon'(q)$ is small. As before, we make a Schrieffer--Wolff transformation on the operators and states, $\cO \to e^A\cO e^{-A}$, $\ket{\Psi} \to e^A\ket{\Psi}$. Anticipating that $A$ will be parametrically small in the ratio $gq/\epsilon'$ (justified \emph{a posteriori}), we have at first order
\begin{equation}
  [A, H_0] + H_1 = 0.
\end{equation}
The solution,
\begin{equation}
  A = \frac{g}{2\sqrt{L}}\sum_{k,q} \frac{\sqrt{kq(q-k)}}{\epsilon(q)-\epsilon(k)-\epsilon(q-k)}\left(\adop_{q}\aop_{q-k}\aop_{k} -\hc \right),
\end{equation}
is indeed small in $gq/\epsilon'$, but has divergent matrix elements between states with a ``soft'' phonon of low--momentum, as anticipated by the vanishing of the quasiparticle residue. Retaining only the singular parts of the generator, we find 
\begin{equation}
  A_\Lambda = \frac{g}{2\sqrt{L}}\sum_{\substack{q \gg \Lambda, \\  0 < k < \Lambda}} \frac{q}{\sqrt{k}\epsilon'(q)}\left(\adop_q\aop_{q-k}\aop_k - \hc\right),
\end{equation}
where $\Lambda$ is some (at the moment arbitrary) parameter less than $q$. 

Another way to think about this is to write the interaction in terms of singular and non--singular parts if we define the hard and soft phonons in terms of their Fourier modes,
\begin{equation}\label{eq:HardSoftModeExpansion}
    \phi^{\gtrless}(x) = -i\frac{1}{\sqrt{L}}\sum_{\substack{q>0\\q\gtrless \Lambda}} \frac{1}{\sqrt{q}}\left(\aop_qe^{iqx} - \adop_qe^{-iqx}\right).
\end{equation}
Then the interaction becomes
\begin{equation}
  \begin{split}
    H_1[\phi_< + \phi_>] &= H_1[\phi_<] + H_1[\phi_>] \\
    &+ \frac{g}{2}\int\dd{x} \partial\phi_<(\partial\phi_>)^2.    
  \end{split}
\end{equation}
The term $\partial\phi_<(\partial\phi_>)^2$ is responsible for the vanishing of the quasiparticle residue, and may be decoupled by $A_\Lambda$ for \emph{any} finite interaction parameter $gq/\epsilon'(q)$. The superficially similar term $\partial\phi_>(\partial\phi_<)^2$ obtained on multiplying out $(\partial\phi_>+\partial\phi_<)^3$ is forbidden by momentum conservation. This leaves behind interactions separated into hard and soft sectors. The hard--hard interactions are necessarily non--singular, but will only be accessible to perturbation theory if also $g\Lambda/\epsilon'(\Lambda)$ is small. Note that disregarding both $H_1[\phi_\gtrless]$ results in the mobile impurity Hamiltonian.

Now we use our generator to calculate the transformed operators that will appear in the correlation function, Eqn.~\eqref{eq:DSF-def}. Acting on a hard phonon, we find
\begin{equation}
    [A_\Lambda, \adop_q] = \frac{g}{\sqrt{L}}\sum_{k<\Lambda}\frac{q}{\sqrt{k}\epsilon'(q)}(\aop_k\adop_{q+k} - \adop_k\adop_{q-k}).
\end{equation}
This can be understood in real space as scaling the hard modes by the soft chiral boson and the dimensionless ratio of interaction to dispersion.
\begin{equation}
  \begin{split}
    [A_\Lambda, \phi^+_>(x)] &= -i\frac{1}{\sqrt{L}}\sum_{q> \Lambda} -i\frac{gq}{\epsilon'(q)}\phi^<(x)\frac{1}{\sqrt{q}} \aop_qe^{iqx}, \\
    [A_\Lambda, \phi^-_>(x)] &= i\frac{1}{\sqrt{L}}\sum_{q>\Lambda} i\frac{gq}{\epsilon'(q)}\phi^<(x)\frac{1}{\sqrt{q}}\adop_qe^{-iqx}.
  \end{split}
\end{equation}
Neglecting the order $\Lambda$ commutator of $\phi^<$ with itself, we can exponentiate the action of $A_\Lambda$ on the hard phonon to find
\begin{equation}\label{eq:RotatedHardBoson}
  \begin{split}
    \phi_>^+(x) &\to -i\frac{1}{\sqrt{L}}\sum_{q>\Lambda}\exp\left(-i\frac{gq}{\epsilon'(q)}\phi^<(x)\right)\frac{1}{\sqrt{q}}\aop_qe^{iqx},  \\
    \phi_>^-(x) &\to \phantom{-}i\frac{1}{\sqrt{L}}\sum_{q>\Lambda}\exp\left(i\frac{gq}{\epsilon'(q)}\phi^<(x)\right)\frac{1}{\sqrt{q}}\adop_qe^{-iqx}.
  \end{split}
\end{equation}
The vacuum remains invariant under the action of $A$, and in the calculation of the structure factor nothing occurs until the vacuum is hit by the density operator $\rho_q$. The phonon then propagates and decays into lower--momentum phonons (thanks to the chiral Hamiltonian).

Continuing to work to first order in $gq/\epsilon'$, the Hamiltonian is free, and therefore upon substituting the rotated hard chiral boson~\eqref{eq:RotatedHardBoson} into the density--density correlation function \eqref{eq:ChiralDecoupledDensityCorrelator}, we find a factorization 
\begin{equation}\label{eq:PhononDecoupledDensityCorrelator}
  \begin{split}
    &\frac{1}{L}\sum_{q>0} \Braket{\exp\left(-i\frac{g_{RL}q}{2u}\phi_L(x,t)\right)\exp\left(i\frac{g_{RL}q}{2u}\phi_L(0,0)\right)}_{H_0}\\
    &\times\Braket{\exp\left(-i\frac{gq}{\epsilon'(q)}\phi_<(x,t)\right)\exp\left(i\frac{gq}{\epsilon'(q)}\phi_<(0,0) \right)}_{H_2^<}\\
    &\times qe^{iqx}\braket{\aop_q(t)\adop_q(0)}_{H_2^>}.
  \end{split}
\end{equation}
The hard correlator produces a factor $e^{-i\epsilon(q)t}$. Using the mode expansion (Eqn.~\eqref{eq:HardSoftModeExpansion}), the soft correlator produces a logarithm, being proportional to
\begin{equation}
  \begin{split}
%    &\Braket{\exp \left(\frac{gq}{\epsilon'(q)}\frac{1}{\sqrt{L}}\sum_{p\ll \Lambda}\frac{1}{\sqrt{p}}\aop_pe^{i[px-\epsilon(p)t]}\right)\exp\left(-\frac{gq}{\epsilon'(q)}\frac{1}{\sqrt{L}}\sum_{p \ll \Lambda}\frac{1}{\sqrt{p}}\adop_p \right)}\\
    &\exp \left(\left(\frac{gq}{\epsilon'(q)}\right)^2\frac{1}{L}\sum_{p\ll \Lambda}\frac{1}{p}e^{i[px-\epsilon(p)t]}\right).
\end{split}
\end{equation}
The summation may be written
\begin{equation}
  \int_{1/L}^{\Lambda}\frac{\dd{p}}{2\pi} \frac{1}{p}e^{i[px-\epsilon(p)t]},
\end{equation}
We put $\Lambda = a q$, with $a\ll 1$ some dimensionless number, and switch to the dimensionless variable $z = px$, in terms of which the summation is  
\begin{equation}
  \int_{x/L}^{aq'x}\frac{\dd{z}}{2\pi z} e^{i[z-\epsilon(z/x)t]} \sim \int_{x/L}^1\frac{\dd{z}}{2\pi z}.
\end{equation}
The integral is dominated by small $z$, and cut off by oscillations of the integrand for $z \sim 1$, leaving just
\begin{equation}
  -\frac{1}{2\pi}\log x/L.
\end{equation}
This leading behaviour misses a factor of $i\pi$, which sends $x \to ix$ and will be crucial in ensuring a real DSF. To see this, we write the sum 
\begin{equation}
  \begin{split}
    \frac{1}{2\pi}\sum_{n=1}^{L \Lambda/2\pi}\frac{1}{n}e^{i 2\pi n x/L} &\sim -\frac{1}{2\pi}\log \left(1-e^{i 2\pi x/L}\right)\\
    &\approx -\frac{1}{2\pi}\log\left( -2\pi i \frac{x}{L}\right),
  \end{split}
\end{equation}
again neglecting the time dependence. 

To obtain the structure factor we must Fourier transform Eqn.~\eqref{eq:PhononDecoupledDensityCorrelator}. As a warm up, we first tackle the chiral case, so that the left--moving factor in Eqn.~\eqref{eq:PhononDecoupledDensityCorrelator} may be dropped and the Fourier transform found exactly, before moving on to the more complicated case accounting for gapless left movers.

\subsection{Chiral problem}

The density--density correlation function obtained in Eqn.~\eqref{eq:PhononDecoupledDensityCorrelator} simplifies to (in the moving frame)
\begin{equation}
  \braket{\rho(x,t)\rho(0,0)} = \int_0^\infty\frac{\dd{q'}}{2\pi} q' (-i[x+i0])^{-\mu(q')}e^{i[q'x-\epsilon(q')t]}.
\end{equation}
Here we have put $\mu(q) = \frac{1}{2\pi}\left(\frac{g_Rq}{\epsilon'(q)}\right)^2$. The infinitesimal shift $x+i0$ ensures that the function is built out of positive wavevectors. A Fourier transform over $t$ gives a delta function on $\omega - \epsilon(q')$, allowing one to carry out the $q'$ integral:
\begin{equation}
  S(x,\omega) = \frac{\epsilon^{-1}(\omega)}{2\pi \epsilon'\left(\epsilon^{-1}(\omega)\right)} (-i[x+i0])^{-\mu(\omega)}e^{i\epsilon^{-1}(\omega) x}.
\end{equation}
Finally, we take a Fourier transform over $x$ to find 
\begin{equation}
  S(q,\omega) = \frac{\epsilon^{-1}(\omega)}{2\pi \epsilon'} \int_{-\infty}^\infty \dd{x} (-i[x+i0])^{-\mu(\omega)}e^{i[\epsilon^{-1}(\omega)-q]x}.
\end{equation}
For $\epsilon^{-1}(\omega) > q$, we may close in the UHP to get zero, and otherwise close around the branch cut on the negative imaginary axis to find the first of Eqn.~\eqref{eq:DSF-chiral-result},
\begin{equation}
  S(q,\omega) = \frac{\epsilon^{-1}(\omega)}{\Gamma\left(\mu(\omega)\right)\epsilon'}\left(q-\epsilon^{-1}(\omega)\right)^{\mu(\omega)-1}\Theta\left(q-\epsilon^{-1}(\omega)\right). 
\end{equation}
As per usual in a leading logarithm calculation, the prefactor should not be taken too seriously.

\subsection{Nonchiral problem}

Life is more complicated if we have to include the left movers. Substituting the power law results into the density--density correlator Eqn.~\eqref{eq:PhononDecoupledDensityCorrelator}, we must Fourier transform 
\begin{equation}
  \begin{split}
    &\int_0^\infty \frac{\dd{q'}}{2\pi} q' (i[x+ut-i0])^{-\mu_{RL}(q')}\\
    &\phantom{\int \dd{q'}}\times(-i[x-ut+i0])^{-\mu_R(q')}e^{i[q'x-\epsilon(q')t]}
  \end{split}
\end{equation}
to obtain the DSF. The infinitesimals ensure that the left--moving factor has negative wavevectors and positive energies, and that the right--moving factor has positive wavevectors and positive energies. We take the transform in sound cone coordinates $x_{L/R} = \frac{1}{2}(x\pm ut)$ to find
\begin{equation}\label{eq:DSF-impurity-integral}
  S(q,\omega) = \int_0^\infty\frac{\dd{q'}}{2\pi} q' \cV_{q'}(q'-q,\omega-\epsilon(q')).
\end{equation} 
Here, $\cV_{q'}(q,\omega)$ is 
\begin{equation}
  \begin{split}
    &\frac{2\pi}{\Gamma(\mu_{RL}(q'))}(\omega-uq)^{\mu_{RL}(q')-1}\Theta(\omega-uq)\\
    &\times \frac{2\pi}{\Gamma(\mu_{R}(q'))}(\omega+uq)^{\mu_R(q')-1}\Theta(\omega+uq).
  \end{split}
\end{equation}
One understands this expression as a sum over all possible hard phonon momenta, which are kinematically constrained by the theta functions. Analytic calculation of the momentum integral is impossible, and we must extract the asymptotics as $\omega \to \epsilon(q)$. We tackle the upward bending phonon dispersion and downward bending phonon dispersion separately.

\subsubsection{Downward bending phonon dispersion}

\begin{figure}[t]
  \centering
  \includegraphics[width=0.8\columnwidth]{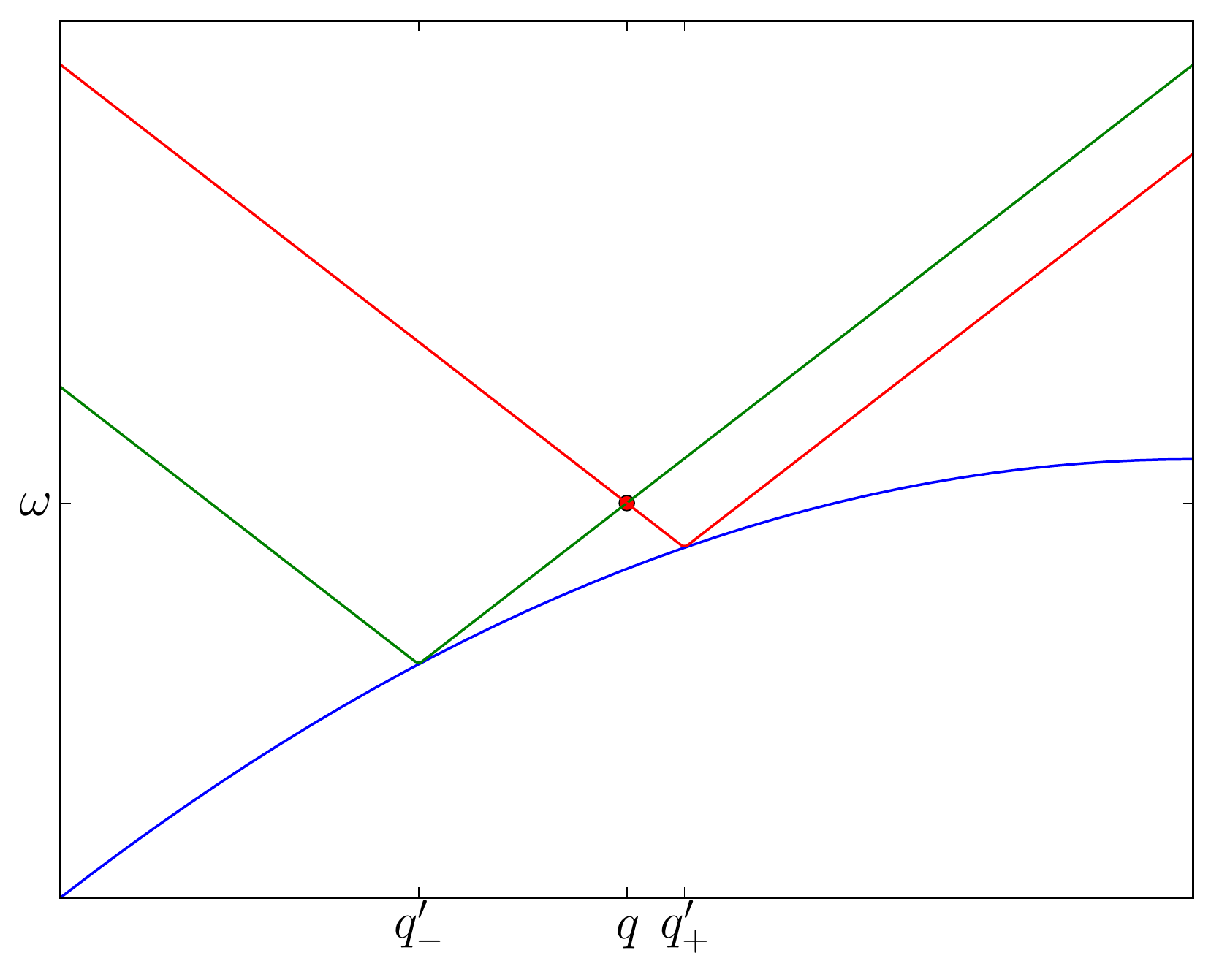}
  \caption{Limits on the possible impurity momenta $q'_\pm$ at given $q, \omega$ (shown by the dot) for the lower threshold.\label{fig:DSF-lower-support}}
\end{figure}

From Fig.~\ref{fig:DSF-lower-support}, we see that as $\omega \to \epsilon(q)$ from above, the range of allowed impurity momenta tends to zero. For $\omega$ below the threshold, the range of integration is zero and the structure factor vanishes. Since the integrand has (integrable) singularities at the upper and lower limits, we cannot treat it as constant, even though the domain of integration is vanishing. We may estimate it by neglecting the variation in the exponents $\mu_R, \mu_{RL}$, and expanding the integrand linearly about the singularities. Thus
\begin{equation}
  S(q,\omega) \sim \int_{q'_-}^{q'_+}\dd{q'}(q'_+-q')^{\mu_{RL}(q)-1}(q'-q_-')^{\mu_R(q)-1}
\end{equation}
and a change of variables brings this to the form
\begin{equation}
(q_+'-q_-')^{\mu_{RL}(q)+\mu_R(q)-1}\int_0^1\dd{s}(1-s)^{\mu_{RL}(q)-1}s^{\mu_R(q)-1}.
\end{equation}
The integral over $s$ is proportional to the Euler Beta function, which is finite in our region of interest where $\mu_R,\mu_{RL}>0$. Since $q'_\pm$ differ from $q$ linearly in $\omega-\epsilon(q)$, we obtain the result for $\mu_2$ in Eqn.~\eqref{eq:NonchiralExponents}, a one--sided power law for the structure factor
\begin{equation}
  S(q,\omega) \sim \left(\omega-\epsilon(q)\right)^{\mu_R(q)+\mu_{RL}(q)-1}\Theta(\omega-\epsilon(q)).
\end{equation}

\subsubsection{Upward bending phonon dispersion}

\begin{figure}[t]
  \centering
  \includegraphics[width=0.8\columnwidth]{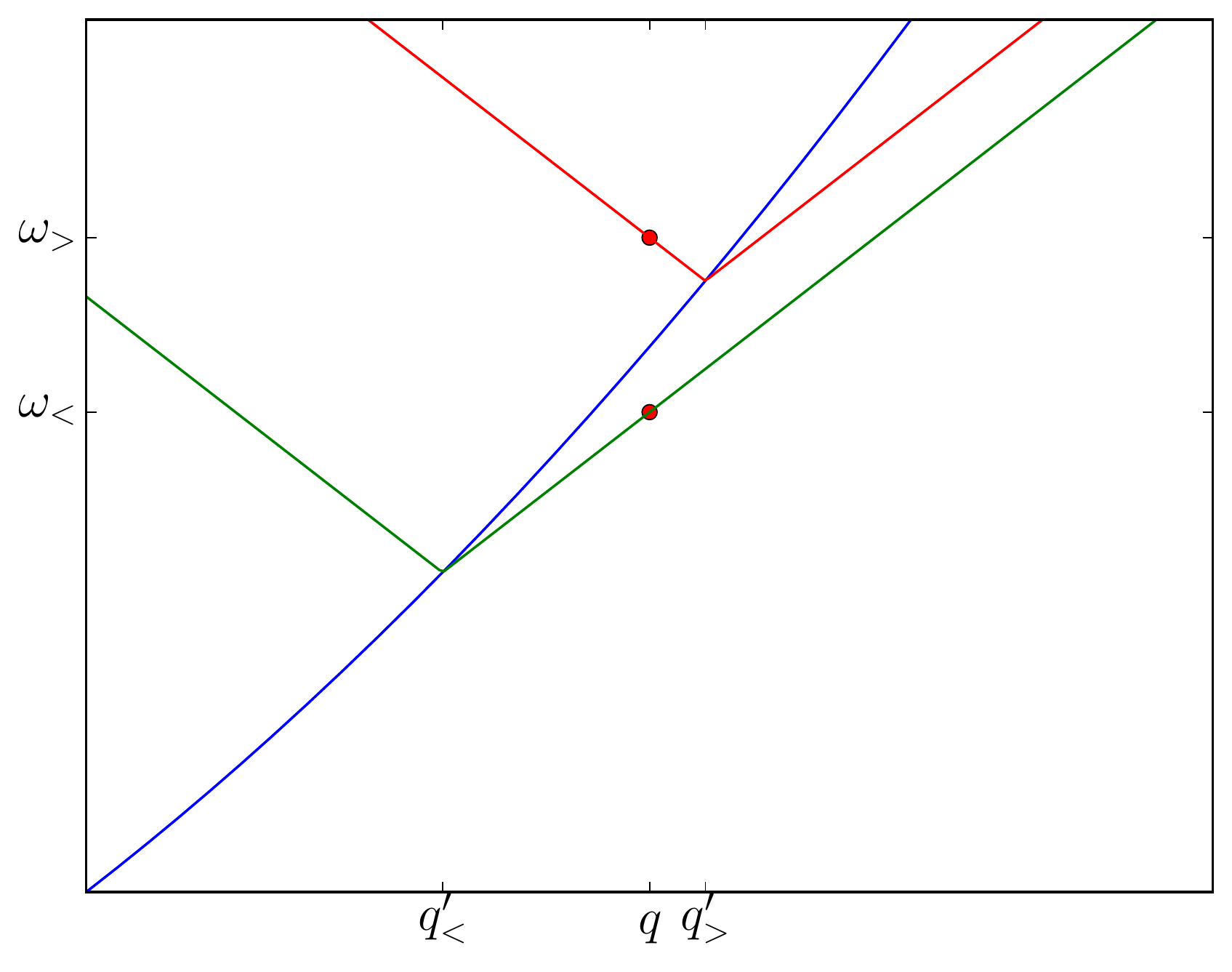}
  \caption{Limits on the possible impurity momenta $q'_\pm$ at given $q, \omega$ (shown by the dots) for the upper threshold.\label{fig:DSF-upper-support}}
\end{figure}

If the dispersion bends upwards, we can see from Fig.~\ref{fig:DSF-upper-support} that the limits on the impurity momenta depend on whether we are above or below the threshold.

For $\omega \lesssim \epsilon(q)$, the integrand has an (integrable) singularity at the upper threshold $q_<'$, corresponding to an impurity plus a bunch of purely right--moving soft excitations that will be responsible for the orthogonality catastrophe. Perhaps perversely, it is the left--handed factor that carries the singularity at $q_<'$, whilst the right--handed factor remains finite as a function of $q'$, but large in the limit $\omega \to \epsilon(q)$. Thanks to the factor $q'/\Gamma(\mu(q'))$, the integrand decreases monotonically to zero at the lower limit, and we can expand the integrand in $q_<'-q'$, where we find
\begin{equation}
  \begin{split}
    S(q,\omega) &\sim \int_0^{q_<'}\dd{q'} (q_<'-q')^{\mu_{RL}(q')-1}(\epsilon(q')-\omega)^{\mu_R(q')-1} \\
    &\sim (\epsilon(q)-\omega)^{\mu_R(q)-1}\int_0^{q_<'}\dd{q'}(q_<'-q')^{\mu_{RL}(q')-1}.
  \end{split}
\end{equation}
The integral is now finite, and the prefactor gives the singularity of the structure factor $S(q,\omega) \sim (\epsilon(q)-\omega)^{\mu_R(q)-1}$ under the threshold.

\begin{figure}[t]
  \centering
  \includegraphics[width=0.8\columnwidth]{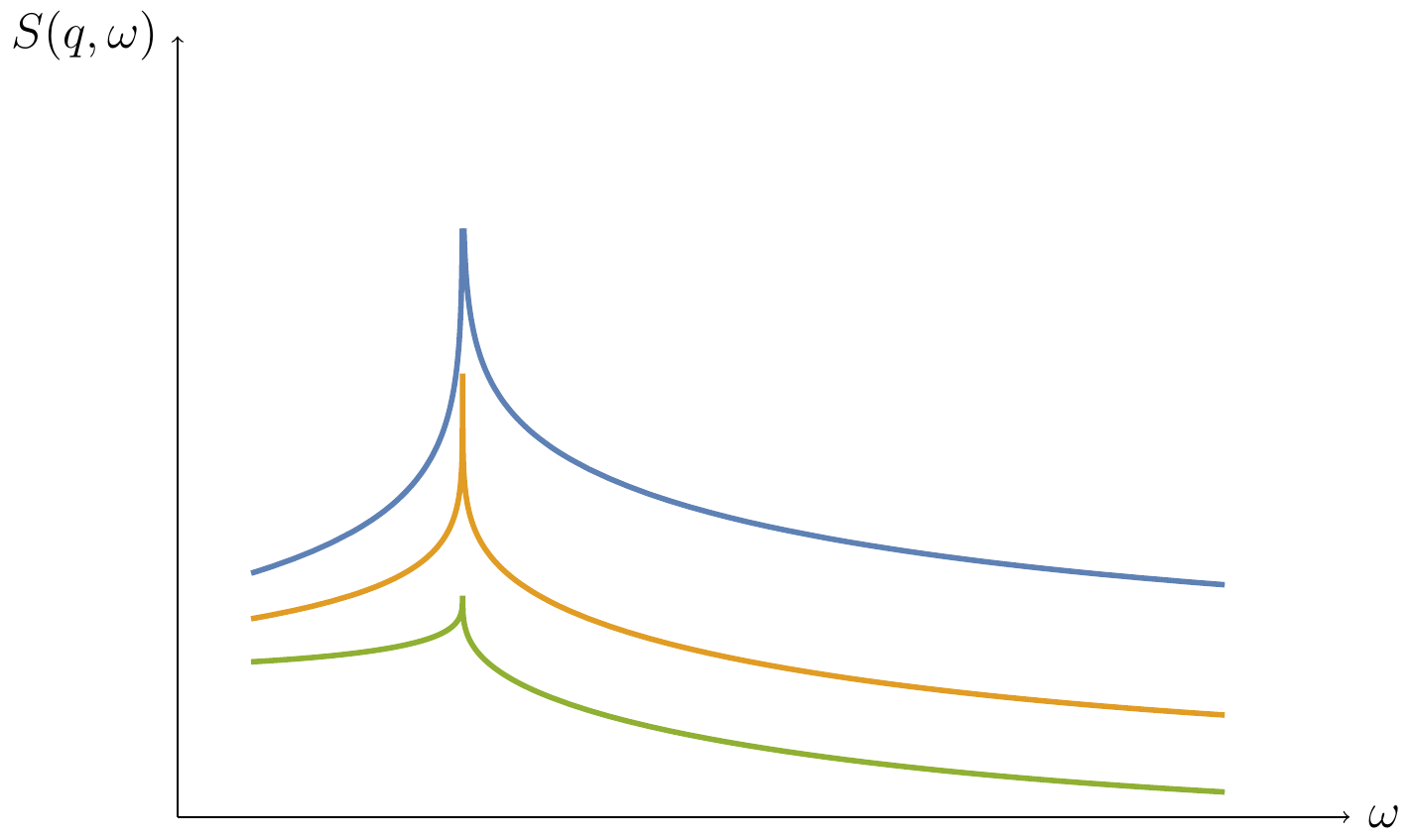}
  \caption{Numerical evaluation of the $S(\omega,q=1)$, Eqn.~\eqref{eq:DSF-impurity-integral}, near the upward--bending Benjamin phonon dispersion, for values of the chiral BO exponent $1/4$ (uppermost curve), $1/2$ (middle curve), and $3/4$ (lowest curve). Note the insensitivity of the super--threshold tail to the BO exponent, as anticipated in Eqn.~\eqref{eq:DSF-super-upward-phonon-asymptote}. \label{fig:numerical-DSF-phonon-upper}}
\end{figure}

For $\omega \gtrsim \epsilon(q)$, the situation is reversed. At the upper threshold, there is a hard phonon plus a bunch of purely left--moving soft excitations, and the right--handed factor carries the singularity at $q_>'$. Playing the same game as before, we expand the structure factor around $q_>'$, where
\begin{equation}\label{eq:DSF-super-upward-phonon-asymptote}
  \begin{split}
    S(q,\omega) &\sim \int_0^{q_>'}\dd{q'}(\omega-\epsilon(q'))^{\mu_{RL}(q')-1}(q_>'-q')^{\mu_{RL}(q')-1}\\
    &\sim (\omega-\epsilon(q))^{\mu_{RL}(q)-1}\int_0^{q_>'}\dd{q'}(q_>'-q')^{\mu_R(q')-1}.
  \end{split}
\end{equation}
This is born out by calculating the integral in Eqn.~\eqref{eq:DSF-impurity-integral} numerically, with results shown for the Benjamin--Ono case in Fig.~\ref{fig:numerical-DSF-phonon-upper}.

These results establish the values of $\mu_3, \mu_4$ in Eqn.~\eqref{eq:NonchiralExponents}, where we have a two--sided singularity with the power laws determined by the orthogonality exponents for the left-- and right--movers respectively.

In summary, the power law singularity at the phonon threshold is given by the logarithmic derivative of the quasiparticle residue,
\begin{equation}
  L\pbyp{Z^{-1}}{L}-1,
\end{equation}
reflecting the orthogonality catastrophe between the ground states of the fluid with and without the hard phonon. 

\subsection{Diagrammatic Calculation}

\begin{figure}[t]
  \centering
  \begin{subfigure}[b]{0.4\columnwidth}
    \includegraphics[width=\textwidth]{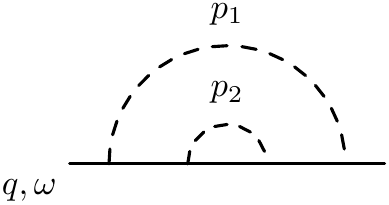}
  \end{subfigure}
  \begin{subfigure}[b]{0.4\columnwidth}
    \includegraphics[width=\textwidth]{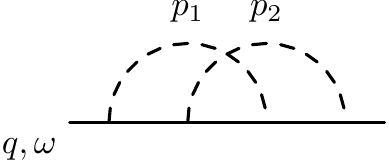}
  \end{subfigure}
  \caption{Corrections to the propagator in the leading logarithm approximation at two loops. Solid lines indicate the ``hard'' phonon propagator, and dashed lines the ``soft'' phonon propagators. \label{fig:SoftTwoLoopDiagrams}}
\end{figure}

Another way to understand the power law in the structure factor that follows from Eqn.~\eqref{eq:PhononDecoupledDensityCorrelator} is by a summation of diagrams. As we noted, the one--loop correction to the self--energy in Fig.~\ref{fig:Self-energy} results in a vanishing quasiparticle residue. We can see this by evaluating it off--shell, where the leading contribution is from soft phonons with momentum $p \ll q$, so
\begin{equation}
  \begin{split}
    \Sigma(\omega, q) &\approx g^2q^2\int_{1/L}^\Lambda\frac{\dd{p}}{2\pi}\frac{p}{\omega-\epsilon(q)+p\epsilon'(q)}\\
    &= \frac{1}{2\pi}\frac{g^2q^2}{\epsilon'(q)} \left[\Lambda + \frac{\omega-\epsilon(q)}{\epsilon'(q)}\log \frac{\omega-\epsilon(q)}{\Lambda \epsilon'(q)} \right]\\
    &\approx [\omega-\epsilon(q)]\frac{1}{2\pi}\left(\frac{gq}{\epsilon'(q)}\right)^2\log \frac{\omega-\epsilon(q)}{q\epsilon'(q)} 
  \end{split}
\end{equation}
for $\omega \sim \epsilon(q)$. In the leading logarithm approximation, we can evaluate higher--order graphs, the first of which at two loops are shown in Fig.~\ref{fig:SoftTwoLoopDiagrams}, and each contributing
\begin{equation}\label{eq:2-loop-leading-log}
  [\omega-\epsilon(q)]\left[\frac{1}{2\pi}\left(\frac{gq}{\epsilon'(q)}\right)^2\log \frac{\omega-\epsilon(q)}{q\epsilon'(q)} \right]^2.
\end{equation}

Diagrams with $n$ loops have the same form as Eqn.~\eqref{eq:2-loop-leading-log}, but with the logarithm and prefactor to the power $n$ instead of 2. It remains to argue that the combinatorial factors conspire to give the $1/n!$ required for the exponential series, which follows from the linked cluster theorem with the single soft loop playing the role of the only disconnected vacuum bubble.

Substituting this into the expression for the propagator $(\omega - \epsilon(q)-\Sigma(\omega,q))^{-1}$, we recover the result Eqn.~\eqref{eq:DSF-chiral-result}. The same leading logarithm summation can equally be found for the ambichiral case.

\section{Soliton Threshold}
\label{sec:Soliton}

In the vicinity of the lower threshold, we calculate the right--moving correlation function $\braket{\aop_q(t)\adop_q(0)}$ appearing in the DSF, Eqn.~\eqref{eq:ChiralDecoupledDensityCorrelator}, with a semiclassical expansion about the soliton, whose density profile was given in Eqn.~\eqref{eq:SolitonSolution}. To justify the semiclassical analysis, consider the Benjamin--Ono example. Rescaling the fields by $\alpha/g$ and lengths by $\alpha^{1/2}$, the action as a function of the rescaled fields becomes
\begin{equation}\label{eq:ChiralAction}
  \left(\frac{\alpha}{g}\right)^2\int \dd{t} \int \dd{x} \left[\partial_x\varphi^-\partial_t\varphi^+ -\frac{1}{2}\partial_x\varphi\partial_x^2\varphi_H - \frac{1}{3!}(\partial_x\varphi)^3\right] 
\end{equation}
and admits a saddle--point analysis when $\alpha/g \gg 1$. However, for notational simplicity, we will carry out the saddle point in the unscaled fields and lengths.

We begin by setting up the apparatus for the chiral boson coherent state path integral. To this end we define the coherent states
\begin{equation}
  \begin{split}
    \ket{\varphi^+} &= \exp \left( i\int \dd{x} \varphi^+(x)\partial\phi^-(x) \right) \ket{0},\\
    \bra{\varphi^-} &= \bra{0}\exp\left(-i\int\dd{x}\partial\phi^+(x)\varphi^-(x)\right),
  \end{split}
\end{equation}
where a $\pm$ superscript indicates the projection onto positive/negative Fourier modes
\begin{equation}
  \begin{split}
    \phi^-(x) &= i\frac{1}{\sqrt{L}}\sum_{k>0}\frac{1}{\sqrt{k}}\adop_ke^{-ikx},\\
    \phi^+(x) &= -i\frac{1}{\sqrt{L}}\sum_{k>0}\frac{1}{\sqrt{k}}\aop_ke^{ikx}.
  \end{split}
\end{equation}
These are the continuum analogue of the analytic coherent state $\ket{\alpha} = \exp{(\alpha \adop)}\ket{0}$. Note that $\phi^\pm(x)$ may be analytically continued into the entire upper/lower half $x$--planes, so $\phi^\pm_H = \pm i\phi^\pm$. The overlap of two coherent states is (c.f. the one--particle case $\braket{\bar{\beta}|\alpha} = e^{\bar{\beta}\alpha}$)
\begin{equation}
  \braket{\varphi^-|\varphi^+} = \exp \left( i\int \dd{x} \partial_x \varphi^-(x)\varphi^+(x)\right).
\end{equation}
Time slicing the density--density correlation function $\braket{0|\aop_qe^{iHt}\adop_q|0}$, and inserting the resolution of the identity
\begin{equation}
  \int \dd{[\varphi^-,\varphi^+]} \exp \left( -i\int \dd{x} \partial_x\varphi^-(x)\varphi^+(x) \right) \ket{\varphi^+}\bra{\varphi^-}
\end{equation}
at each slice we find a path integral representation for the propagator $K(\varphi_F^-,\varphi_I^+,t) = \braket{\varphi_F^-|e^{-iHt}|\varphi_I^+}$, 
\begin{equation}
  \int\displaylimits_{\substack{\varphi^+(0) =\varphi_I^+\\ \varphi^-(t) = \varphi_F^-}} \hspace{-5mm}\cD(\varphi^-,\varphi^+) \exp\left(-i\int\dd{x}\partial_x \varphi_F^-\varphi^+(t) + iS[\varphi^-,\varphi^+] \right),
\end{equation}
where $S$ is the action defined in Eqn.~\eqref{eq:ChiralAction}. Note that we specify only $\varphi^+$ at $t=0$ and $\varphi^-$ at $t$; to specify the full function $\varphi$ would overdetermine the problem, akin to fixing both $p$ and $q$ at the beginning and end of a problem in classical mechanics. The boundary term $\int \dd{x\,} \partial_x\varphi^-_F\varphi^+(t)$ appears on a careful analysis of the time--slicing (see Ref.~\onlinecite{Faddeev:1976}, and Appendix~\ref{appendix:BoundaryTerm} for an explicit demonstration). It ensures that the semiclassical expression for the propagator retains the correct analytic dependence on $\varphi^+_F, \varphi^-_I$,
\begin{equation}
  \begin{split}
    &\phantom{=}\delta  \log K^{\text{cl}}(\varphi^-_F,\varphi^+_I,t) \\
    &= -i \int \dd{x} \delta [\partial_x\varphi_F^-] \varphi^+(t) + i \int \dd{x} \partial_x\varphi^-(0) \delta \varphi^+_I,
  \end{split}
\end{equation}
which amounts to a derivation of the Hamilton--Jacobi equations. Despite its smallness compared to the ``bulk'' action, the boundary term will have a crucial role to play in coupling the soliton to Gaussian fluctuations.

We use the propagator to take care of the time evolution in the density--density correlator $q\braket{\aop_q(t)\adop_q(0)}$, which becomes
\begin{equation}
  \begin{split}
    &q\int \dd{[\varphi_F]}\dd{[\varphi_I]}e^{-\int \partial_x\varphi_F^-\varphi^+_F}e^{-\int\partial_x\varphi_I^-\varphi^+_I}\\
    &\times\braket{0|\aop_q|\varphi^+_F}K(\varphi^+_F,\varphi^-_I,t)\braket{\varphi^-_I|\adop_q|0}.
  \end{split}
\end{equation}
The matrix elements of the density operator are $[\partial_x\varphi^+_F]_q$ and $[\partial_x\varphi^-_I]_q$ respectively. We shift our integration field $\varphi \to \Phi + \tilde{\varphi}$, where $g\partial_x\Phi = V$, the classical soliton solution of Eqn.~\eqref{eq:SolitonSolution}. We then drop the $\tilde{\varphi}$ terms outside the exponential, where they will provide small additive corrections to the final result, so that the correlator becomes
\begin{equation}
  \int \cD{(\tilde{\varphi}^-,\tilde{\varphi}^+)} [\partial\Phi_F^+]_q[\partial\Phi_I^-]_qe^{i[S_0+S_1 + S_2]}.
\end{equation}
The linear term in $S_1$ results from the boundary terms discussed previously 
\begin{equation}\label{eq:ActionLinearVariation}
  S_1 = \int\dd{x}\partial_x\Phi^-(x,t)\tilde{\varphi}^+(x,t) -\int\dd{x}\partial_x\Phi^+(x,0)\tilde{\varphi}^-(x,0)
\end{equation}
and is the chiral boson analogue of the single--particle result Eqn.~\eqref{eq:EndPointVariation}. The linear bulk variation vanishes by the equations of motion, leaving the quadratic term
\begin{equation}\label{eq:ActionQuadraticVariation}
  \begin{split}
    S_2 &= -\int\dd{x}\partial_x\tilde{\varphi}^-(0)\tilde{\varphi}_I^+\\
    &\phantom{=}+\int_0^t \dd{t}\dd{x}\left[\partial_x\tilde{\varphi}^-\partial_t\tilde{\varphi}^+ -\cH_0(\tilde{\varphi}) - \frac{g}{2}\partial_x\Phi(\partial_x\tilde{\varphi})^2\right]
  \end{split}
\end{equation}
controlling the Gaussian fluctuations of $\tilde{\varphi}$ about the soliton.

At this point, we can step back and observe that due to the localized profile of the soliton, the integration over the linear variation in Eq.~\eqref{eq:ActionLinearVariation} is essentially $i\Delta [\tilde{\varphi}^+(X(t))-\tilde{\varphi}^-(X(0)]$, where $\Delta$ is the soliton weight, so that the correlation function amounts to that of a vertex operator in a Gaussian action. The resulting power law is accompanied by a factor $e^{-iE_St}$ from the zeroth--order evolution of the soliton, so that in Fourier space the singularity sits on the soliton dispersion, which anticipates the second of our results for the chiral correlation function in Eqn.~\eqref{eq:DSF-chiral-result}. However, we must address the fact that the quadratic action \eqref{eq:ActionQuadraticVariation} is not a Luttinger Liquid, and must find a basis diagonalizing it to calculate the Gaussian integral. In both the KdV and BO cases, its eigenvalues are identical with and without the soliton. This is presumably due to the integrability of each model, however, this detail does not affect the asymptotics of the final result, which arises from the linear dispersion of the phonons $v_Sk$ seen from the frame of the soliton. Further details of these calculations differ in the KdV and BO cases, which we tackle in turn.

\subsection{Benjamin--Ono}

Using the expression for $\partial_x \Phi$ as a sum over poles, 
\begin{equation}\label{eq:BO-soliton-pole-form}
  \partial_x\Phi(x,t) = -\frac{2i\alpha}{g} \left[\frac{1}{x-X(t)} -\frac{1}{x-\bar{X}(t)}\right]
\end{equation}
the phonon--soliton coupling term Eqn.~\eqref{eq:ActionLinearVariation} may be evaluated exactly as
\begin{equation}
  \frac{4\pi i\alpha}{g} \left[\tilde{\varphi}^+(X(t),t) - \tilde{\varphi}^-(\bar{X}(0),0) \right],
\end{equation}
where the prefactor is indeed the soliton weight $\Delta$, as anticipated. Taking a Fourier transform of the soliton, Eqn.~\eqref{eq:BO-soliton-pole-form}, we find the matrix elements $\partial_x\Phi^\pm(q)$ appearing in the correlation function:
\begin{equation}
  \begin{split}
    [\partial_x\Phi^+_F ]_q &= \frac{4\pi \alpha}{g} e^{iqX_F},\\
    [\partial_x\Phi^-_I]_q &= \frac{4\pi \alpha}{g}e^{-iq\bar{X}_I}.
  \end{split}
\end{equation}
To diagonalize the quadratic action, Eqn.~\eqref{eq:ActionQuadraticVariation}, we use the basis found in Ref.~\onlinecite{Chen:1980}
\begin{equation}
  \begin{split}
    \psi^+_k(y) &= \frac{y-X}{y-\bar{X}}\left[\frac{1}{i(k+v_S/2\alpha)(y-X)}-1\right]e^{iky},\\
    \psi^-_k(y) &= \frac{y-\bar{X}}{y-X}\left[\frac{i}{(k+v_S/2\alpha)(y-\bar{X})} -1\right]e^{-iky},
  \end{split}
\end{equation}
where $y= x-v_St$. These solutions scatter with neither reflection or phase shift (mod $2\pi$). We expand the phonon field on this basis as
\begin{equation}
  \begin{split}
    \tilde{\varphi}^+(y,t) &= \int_0^\infty\frac{\dd{k}}{2\pi} \eta_k(t)\psi^+_k(y),\\
    \tilde{\varphi}^-(y,t) &= \int_0^\infty\frac{\dd{k}}{2\pi} \bar{\eta}_k(t)\psi^-_k(y).
  \end{split}
\end{equation}
The normalization is fixed by continuity with the $y\to \infty$ limit, where $\psi^\pm_k \to e^{\pm iky}$. Then a calculation of the projector kernel
\begin{equation}
  \int_0^\infty \frac{\dd{k}}{2\pi}\psi^+_k(y)\psi^-_k(y') 
\end{equation}
gives a correctly normalized delta function, as it should. The quadratic action is diagonal by construction, and in fact coincides with the non--interacting case
\begin{equation}
  \int_0^t{\dd{t'}}\int_0^\infty \frac{\dd{k}}{2\pi} \bar{\eta}_k(t')k\underbrace{[i\partial_{t'}-\epsilon(k)]}_{G_0^{-1}(k,t')}\eta_k(t').
\end{equation} 
``Completing the square'' in the usual way, we carry out the integration over the $\eta_k$ to obtain
\begin{equation}
  \begin{split}
    &\phantom{=}\Delta^2\int_0^\infty\frac{\dd{k}}{2\pi} \frac{1}{k} \psi^-_k(X(t)-v_St,t)G_0(k,t)\psi^+_k(\bar{X}(0),0).\\
%    &= \Delta^2\int_0^\infty\frac{\dd{k}}{2\pi}\frac{1}{k} e^{i[k(X(t)-\bar{X}(0)-v_St)-\epsilon(k)t]}\psi_k^-(y,t)\psi_k^+(0,0)\\
  \end{split}
\end{equation}
The Green function evolves the phonons with a phase $e^{-i\epsilon(k)t}$. For $v_st \gg l_S$, the resulting integral over $k$ is dominated by momenta much smaller than the inverse soliton length. $\psi^-_k(X(t)-v_St)$ tends to $-e^{-ik[X(t)-v_St]}$, whilst $\psi_k^+(X(0),0)$ tends to 
\begin{equation}
  \frac{1}{i(v_S/2\alpha)[\bar{X}(0)-X(0)]}e^{ikX(0)}.
\end{equation}
Since $\bar{X}-X = 2l_S = 2 \alpha/v_S$, this reduces to $-ie^{ikX(0)}$. Altogether, the integration over Gaussian fluctuations is
\begin{equation}
  \exp \left(\frac{\Delta^2}{2\pi}\int_{1/L}^{1/l_S}\frac{\dd{k}}{k}e^{ik[v_St-X(0)-\bar{X}(0)]-i\epsilon(k)t}\right).
\end{equation}
The dominant phase at low $k$ and $v_St \gg l_S$ is simply the $e^{ikv_St}$, giving a factor
\begin{equation}
  (v_St/L)^{-\Delta^2/2\pi}.
\end{equation}
The soliton weight that appears in the exponent is to be determined by the zero--mode integration, to be discussed after consideration of the KdV case, that fixes the momentum of the soliton, and hence its weight, energy, and velocity via the classical equations of motion.

\subsection{Korteweg--de Vries}

The KdV soliton may be written as a sum over infinitely many double poles
\begin{equation}\label{eq:KdV-soliton-pole-form}
  \partial_x\Phi^\mp(x,t) = \frac{3iv_S}{g}\sum_{n\gtrless 0} \frac{1}{[x-X(t)- X_n]^2},
\end{equation}
where $X_n = i(n+1/2)\pi l_S$, and $l_S = \frac{1}{2}\sqrt{v_S/\beta}$ is the soliton length. The phonon--soliton coupling, Eqn.~\eqref{eq:ActionLinearVariation}, is
\begin{equation}
  \frac{24 \pi \beta}{g}\sum_{n=0}^\infty \partial\tilde{\varphi}^+(X_n,t),
\end{equation}
which is not immediately recognizable as $i\Delta \tilde{\varphi}^+(X)$. That it is so can be seen by Fourier expanding $\tilde{\varphi}$, so that the coupling is
\begin{equation}
  \begin{split}
    &\phantom{=}\frac{24\pi i \beta}{g} \int_0^\infty\frac{\dd{k}}{2\pi} k \tilde{\varphi}_k(t)e^{ikX}\sum_{n=0}^\infty e^{ikX_n}\\
    &=\frac{24\pi i \beta}{g} \int_0^\infty \frac{\dd{k}}{2\pi} \frac{k}{2\sinh(\pi l_Sk/2)} \tilde{\varphi}_ke^{ikX}.
  \end{split}
\end{equation}
For functions that are flat on the scale of the soliton length, this simplifies to
\begin{equation}
  \frac{24i\beta}{g l_S} \tilde{\varphi}(X) = \frac{12i\sqrt{\beta v_S}}{g}\tilde{\varphi}(X),
\end{equation}
which is indeed $i\Delta \tilde{\varphi}(X)$.

We find the matrix elements $[\partial_x\Phi^\pm]_q$ by Fourier transforming the soliton, Eqn.~\eqref{eq:KdV-soliton-pole-form},
\begin{equation}
  \begin{split}
    [\partial_x\Phi^{\mp}]_q &= \frac{12\beta\cdot 2\pi}{g}\sum_{n\gtrless 0} qe^{iq(X-X_n)}\\
    &= \frac{12\pi\beta q}{g}\frac{e^{iqX(t)}}{\sinh \pi l_Sq/2}.
  \end{split}
\end{equation}
For $ql_S \ll 1$, the prefactor simplifies to the inverse soliton length, leaving only the important $e^{iqX(t)}$ dependence. 

The solutions to the linear scattering problem $\cA e^{iky}$ are conveniently expressed in terms of the ``intertwiner'' \cite{Koller:2015}
\begin{equation}
  \cA = \partial^4 - \frac{l_Sv_S}{\beta}\tanh \frac{y}{l_S} \partial^3 + \frac{2}{3\beta}V\partial^2 - \frac{1}{9\beta}V'\partial,
\end{equation}
which has the defining property $\cA H_0 = H_1\cA$, where $H_0$ is the quadratic action without the soliton, and $H_1$ is the quadratic action in the presence of the soliton. Given a solution $H_0e^{iky} = \epsilon(k)e^{iky}$, it follows that $\cA e^{iky}$ is an eigenvector of $H_1$ with the same eigenvalue $\epsilon(k)$. As before, we normalize $\cA e^{iky}$ by demanding that it tends to the free solution $e^{iky}$ far from the soliton, which is achieved by dividing out the leading factor $k^4$. Now we expand our fluctuations $\tilde{\varphi}$ on these scattering states:
\begin{equation}
  \tilde{\varphi}(x,t) = \int_0^\infty\frac{\dd{k}}{2\pi}\eta_k(t)\cA e^{iky},
\end{equation}
and complete the square on the Gaussian integral to find
\begin{equation}
  \begin{split}
    &\sum_{n,m > 0}\left(\frac{24\pi \beta}{g}\right)^2\int_0^\infty\frac{\dd{k}}{2\pi}\frac{1}{k}e^{-i\epsilon(k)t}\\
    &\phantom{\sum}\times[\partial_y\cA e^{iky}]_{y = X(t) + X_n}[\partial_y\cA e^{-iky}]_{y = X(0) -X_m}.
  \end{split}
\end{equation}
In the limit $v_St \gg l_S$, $l_Sk \ll 1$, the scattering solutions are dominated by their leading terms $e^{iky}$. Summing this over the locations of poles, the Gaussian integral is asymptotically
\begin{equation}
  \exp \left(\frac{\Delta^2}{2\pi} \int_{1/L}^{1/l_S}\frac{\dd{k}}{k} e^{ikv_St}\right), 
\end{equation}
giving a factor $(v_St/L)^{-\Delta^2/2\pi}$ exactly as in the BO case. The difference is in the momentum--dependence of the soliton weight:
\begin{equation}
  \Delta = \int \dd{x} \partial_x\Phi = \left( \frac{12^2\beta P}{g}\right)^{1/3},
\end{equation}
where $P$ is the soliton momentum.

\subsection{Zero modes}

Varying $X(t), \bar{X}(t)$, the bulk action is stationary provided the soliton obeys the classical equations of motion. The condition for the vanishing boundary variation is
\begin{equation}
  0 = \delta X_F\left[q - \int \dd{x}\partial_x\Phi^-\partial_x\Phi^+\right],
\end{equation}
which simply says that the momentum of the soliton must be equal to $q$. Since the solution obeys the classical equations of motion, the bulk action causes it to evolve with a phase $e^{-iE_S(q)t}$. There are no $t$-- dependent corrections to this result from the Gaussian integral over fluctuations in $X,\bar{X}$ since the calculation is analogous to the single--particle free propagator between momentum eigenstates, whose time dependence is entirely given by the phase.

The result of the semiclassical calculation is that the chiral correlation function $q\braket{\aop_q(t)\adop_q(0)}$ is asymptotically
\begin{equation}\label{eq:SolitonCorrelator}
  t^{-\Delta(q)^2/2\pi}e^{-iE_S(q)t}.
\end{equation}
To evaluate the structure factor, we substitute our result Eqn.~\eqref{eq:SolitonCorrelator} into the chiral--decoupled density--density correlation function \eqref{eq:ChiralDecoupledDensityCorrelator}. If the left--movers are gapped out, the Fourier transform proceeds as in the phonon case, and we find the second  and final of our main result Eqn.~\eqref{eq:DSF-chiral-result} for the gapped left--movers,
\begin{equation}
  S(q,\omega) \sim (\epsilon^{-1}(\omega)-q)^{\Delta(q)^2/2\pi-1}\Theta(\epsilon^{-1}(\omega)-q).
\end{equation}
Since the soliton weight is large, the structure factor vanishes as a power law at the threshold.

If we have gapless left--movers, we must consider the case of upward--bending and downward--bending soliton dispersion in turn.

\subsubsection{Downward bending soliton dispersion}

As for the phonons, the domain of integration vanishes, and we confirm our final result for $\mu_1$ in Eqn.~\eqref{eq:NonchiralExponents},
\begin{equation}
  S(q,\omega) \sim (\omega-E(q))^{\Delta(q)^2/2\pi + \mu_{RL}(q)-1}.
\end{equation}
The resulting power law suppression of the structure factor is largely unaffected by the left--movers since $\Delta^2 \gg 1 \gg \mu_{RL}$.

\subsubsection{Upward bending soliton dispersion}

Putting $\omega = E(q)$ and $\mu_R = \Delta^2/2\pi \gg 1$ into the integral for the structure factor Eqn.~\eqref{eq:DSF-impurity-integral}, the integral is convergent, and the structure factor remains finite. There is no trace of the soliton in the structure factor at the upper threshold.
\section{Conclusions}
In summary, the linear Luttinger liquid theory fails to resolve the frequency dependence of correlation functions, a consequence of having dispensed with the energy scales of phonon interaction and dispersion. We have shown that it is possible to resolve the singular behaviour of correlation functions from the nonlinear hydrodynamics, without requiring the introduction of a mobile impurity model at the outset. Our results are valid at momenta much smaller than $u/g_{RL}$, but large enough that the chiral dispersive effects dominate the interaction. In the semiclassical regime, there are singularities near the mass shells of the classical excitations (sound waves and solitons). The fractional parts of the exponents are given by the power law governing the Anderson orthogonality catastrophe for the phonons, and the soliton weight, respectively. Whilst these results were obtained in the semiclassical limit, the physical mechanism (the orthogonality catastrophe) does not rely on this approximation, even though steps in the calculation do, and therefore we expect these results to remain valid beyond the parameter range where they strictly apply.

It is interesting to wonder which of the (chiral) cases discussed here, if any, are relevant to quantum hall edge states.  As well as the implications for the structure factor and spectral function, one may wonder whether there are any consequences for the thermalization of these states.  The effect of phonon dispersion on thermalization of fermions, meaning the electron distribution function, has been discussed in Ref.~\onlinecite{Kovrizhin:2011}. Conversely, phonon interactions allow the phonon distribution function $n_k$ to change with time (unlike the linear LL), so that one might naively expect the phonon distribution function of an isolated nonlinear LL to equilibrate. However, a full calculation remains to be done, and the question is subtle in the strongly quantum limit of vanishing dispersion, where there is no perturbation theory for the phonons and one must instead calculate correlation functions in terms of free fermions.

Our calculations are strictly for the zero--temperature correlation functions. In practice, one expects these results to be relevant when the temperature is much less than that of the energy scales $\epsilon(k), E(k)$ of the hard phonons and solitons. An obvious but challenging extension of this work, and related to the above question on thermalization, would be to find the finite--temperature correlation functions.

We have focused here only on the hydrodynamics of the density. In the linear Luttinger liquid theory, and in certain integrable models, spin and charge degrees of freedom separate.  Work has begun on reconciling this with nonlinear Luttinger liquid phenomenology\cite{Essler:2015}, but there is much to understand. Whether one can use a nonlinear two component hydrodynamic theory describing the spin and charge to calculate correlation functions (which may be benchmarked against the known results for integrable models), is a difficult open problem.

\emph{Acknowledgment} T.P. gratefully acknowledges support of the EPSRC.

\bibliography{../Literature/QHD}

\appendix

\section{Refermionization of linearly--dispersing phonons}\label{Appendix:Refermionization}
Here we summarize the arguments that bosonization of free chiral fermions with band curvature gives linearly--dispersing chiral phonons with cubic interactions. We will do so in first quantization, following Refs.~\onlinecite{Stone:1994, stone2008quantum}, and write our fermionic wavefunctions as
\begin{equation}
  \psi(\{z_i\}) = \Phi(\{z_i\}) \Delta.
\end{equation}
We work on a ring with coordinates $z_i = e^{i\theta_i}$. Here, $\Phi$ is a symmetric function encoding excitations over the fermionic ground state $\Delta =\det z_i^p$, a Slater determinant. The Hamiltonian for free fermions is
\begin{equation}
  H_F = \frac{1}{2m}\sum_i\left(z_i\partial_i\right)^2.
\end{equation}
 To bosonize the system, we wish to write matrix elements of the fermionic system $\braket{\psi_1|\psi_2}_F$ as a function of the power sums $p_n = \sum_i z_i^n$. The $p_n$ are nothing but the Fourier modes of the density $\sum_i\delta(z-z_i)$. It can be shown \cite{Stone:1994} that the change of variables in the measure is
\begin{equation}
  \prod_i\frac{\dd{\theta_i}}{2\pi} |\Delta|^2 \to \prod_n \frac{\dd{\bar{p}_n}\dd{p_n}}{\pi n} \exp \left(-\sum_{n>0}\frac{1}{n}\bar{p}_np_n\right).
\end{equation}
Given this inner product, the adjoint is given by $p_n^\dagger = n\pbyp{}{p_n}$, which results in the Kac--Moody algebra $[p_n, p_m^\dagger] = n \delta_{n+m,0}$ familiar from the second--quantized result. It is then an exercise in calculus to show that $\braket{\Phi_2\Delta|H_F|\Phi_1\Delta}_F = \braket{\Phi_2|H_B|\Phi_1}$, where  
\begin{equation}
  H_B = \frac{1}{2m}\sum_{r,s>0}rsp_{r+s}\pbyp{}{p_{r}}\pbyp{}{p_s} + (r+s)p_rp_s\pbyp{}{p_{r+s}},
\end{equation}
plus a piece proportional to the linear chiral Luttinger Liquid, that we can remove by Galilean transformation. Comparing with Eqn.~\eqref{eq:MomentumSpaceInteraction}, this is a chiral coupling between phonons with $g_R = 1/m$, and $g_{RL}=0$. In the absence of $2k_F$ processes, a nonchiral phonon coupling arises from a density--density interaction $\partial\phi_L\partial\phi_R$. The quadratic part is diagonalized by a unitary transformation on the $\phi_L, \phi_R \to \tilde{\phi}_L, \tilde{\phi}_R$, in terms of which the cubic term is no longer chiral.

\section{Boundary Term in the Coherent State Path Integral from Time Slicing}\label{appendix:BoundaryTerm}
Here we derive the boundary term for the single--particle coherent state path integral, since this is notationally cleaner and therefore easier to follow than the chiral boson used in the text (the generalization to which should be clear). This material is far from original, but is often missed in textbook discussions (but see Ref.~\onlinecite{Faddeev:1976}).
\subsection{Propagator}
Slice the propagator $K(\bar{\alpha}_F,\alpha_I, t) = \braket{\bar{\alpha}_F|e^{-iHt}|\alpha_I}$ into $N$ intervals of duration $\Delta t = t/N$,
\begin{equation}
  \bra{\bar{\alpha}_F}e^{-iH\Delta t}_{\phantom{-iH\Delta t}\bigwedge_{N-1}} e^{-iH\Delta t}_{\phantom{-iH\Delta t}\bigwedge_{N-2}} \cdots \phantom{e}^{\phantom{-iH\Delta t}}_{\phantom{-iH\Delta t}\bigwedge_1} e^{-iH\Delta t}\ket{\alpha_I},
\end{equation}
and insert a resolution of the identity
\begin{equation}\label{eq:CoherentStateResolutionIdentity}
  \int \dd{[\bar{\alpha},\alpha]} e^{-\bar{\alpha}\alpha}\ket{\alpha}\bra{\bar{\alpha}}
\end{equation}
at each of the $N-1$ points indicated by $\wedge_j$. At $\wedge_j$, one obtains a factor of $e^{-iH(\bar{\alpha}_{j+1},\alpha_j)\Delta t}\braket{\bar{\alpha}_{j+1}|\alpha_j}$ from the propagator. The coherent state overlap $\braket{\bar{\alpha}_{j+1}|\alpha_j} = e^{\bar{\alpha}_{j+1}\alpha_j}$ combines with the factor $e^{-\bar{\alpha}_{j+1}\alpha_{j+1}}$ arising from the normalization factor in the resolution of the identity to provide the exponential of
\begin{equation}\label{eq:CoherentStateOverlaps}
  \bar{\alpha}_F\alpha_{N-1}-\bar{\alpha}_{N-1}\alpha_{N-1} + \bar{\alpha}_{N-1}\alpha_{N-2} - \cdots - \bar{\alpha}_1\alpha_1 + \bar{\alpha}_1\alpha_I. 
\end{equation}
Rearranging this and formally taking the limit we arrive at the usual phase space term in the action, plus a boundary term
\begin{equation}
  \begin{split}
    & \bar{\alpha}_F\alpha_{N-1} - \sum_{\substack{n=1\\\alpha_0 = \alpha_I}}^{N-1}\hspace{-1mm} \bar{\alpha}_n(\alpha_n-\alpha_{n-1}) \to \bar{\alpha}_F\alpha(t) - \int_0^t \dd{t}\bar{\alpha}\partial_t\alpha.
  \end{split}
\end{equation}
In the above, $\alpha(0) = \alpha_I$, and the boundary term $\bar{\alpha}_F\alpha(t)$ is left over as claimed.
\subsection{Two--point correlator}
The object required in the text is the analogue of $\braket{0|\aop(t)\adop(0)|0}$. This is obtained from the propagator by
\begin{equation}
  \int \dd{(\bar{\alpha}_F,\alpha_F)}\dd{(\bar{\alpha}_I,\alpha_I)} e^{-\bar{\alpha}_F\alpha_F}e^{-\bar{\alpha}_I\alpha_I}\alpha_FK(\bar{\alpha}_F,\alpha_I,t)\bar{\alpha}_I.
\end{equation} 
Using the explicit form of the time--sliced propagator, the overlap of the coherent states Eqn.~\eqref{eq:CoherentStateOverlaps} gains the two terms $-\bar{\alpha}_F\alpha_F- \bar{\alpha}_I\alpha_I$ to become
\begin{equation}
  \begin{split}
    &\phantom{=}-\bar{\alpha}_F\alpha_F  + \bar{\alpha}_F\alpha_{N-1} - \sum_{\substack{n=1\\\alpha_0 = \alpha_I}}^{N-1}\bar{\alpha}_n(\alpha_{n}-\alpha_{n-1}) - \bar{\alpha}_I\alpha_I\\
    &= -\hspace{-3mm}\sum_{\substack{n=0\\ \bar{\alpha}_N=\bar{\alpha}_F, \\ \alpha_0 = \alpha_I}}^{N}\hspace{-3mm}\bar{\alpha}_n(\alpha_n-\alpha_{n-1}) - \bar{\alpha}_I\alpha_I \to -\int_0^t\dd{t}\bar{\alpha}\partial_t\alpha - \bar{\alpha}_I\alpha_I.
  \end{split}
\end{equation}
This expression leads to the endpoint variation
\begin{equation}\label{eq:EndPointVariation}
  -\bar{\alpha}(t)\delta\alpha(t) - \delta\bar{\alpha}(0)\alpha(0)
\end{equation}
of Eqn.~\eqref{eq:ActionLinearVariation}.

\end{document}